\documentclass[12pt,preprint]{aastex}

\newcommand{\pks}{PKS~2155-304}
\newcommand{\as}{S\'{e}rsic~159-03}
\newcommand{\xmm}{{\it XMM-Newton}}
\newcommand{\rosat}{{\it ROSAT}}

\begin{document}

\title{Thermal and non-thermal nature of the soft excess emission from \as\/
observed with \xmm
}

\author{Massimiliano~Bonamente$\,^{1,2}$, Richard~Lieu$\,^{2}$, Jonathan~P.~D.~Mittaz$\,^{2}$, 
Jelle~S.~Kaastra$\,^{3}$ and Jukka~Nevalainen$\,^{4}$
}

\affil{
\(^{\scriptstyle 1} \)
{NASA National Space Science and Technology Center, Huntsville, AL}\\
\(^{\scriptstyle 2} \)
{Department of Physics, University of Alabama, Huntsville, AL}\\
\(^{\scriptstyle 3} \)
{SRON Utrecht, The Netherlands}\\
\(^{\scriptstyle 4} \)
{Observatory, University of Helsinki, Finland}\\
}

\begin{abstract}
\noindent~Several nearby clusters exhibit an excess of soft X-ray radiation which
cannot be attributed to the hot virialized intra-cluster medium. There is no
consensus to date on the origin of the excess emission: it could be either
of thermal origin, or due to an inverse Compton scattering of the cosmic
microwave background. 
Using high resolution \xmm~ data of \as\/
we first show that strong soft excess emission is detected out to a radial distance
of 0.9 Mpc.
The data are interpreted using the two viable models available, i.e., by
invoking a warm reservoir of thermal gas, or relativistic electrons
which are part of a cosmic ray population.  
%The thermal model leads to a better goodness-of-fit, and the emitting warm gas
%must be high in mass and low in metallicity.
The thermal interpretation of the excess emission, slightly favored by the
goodness-of-fit analysis, indicates that the warm gas responsible for the 
emission is high in mass and low in metallicity.
\end{abstract}


\section{Introduction}

Galaxy clusters often feature an excess of soft X-ray radiation above the contribution from the
hot intra-cluster medium (ICM), known as the {\it cluster soft excess} phenomenon. 
Among the clusters that display this phenomenon is \as,
whose soft excess was initially discovered from \rosat~PSPC data by 
Bonamente, Lieu and Mittaz (2001),
and later confirmed by an earlier \xmm~ observation of Kaastra et al. (2003);
the \as~ excess is one of the strongest observed to date. 
In this paper we analyze a deep (122 ks) \xmm~ observation of this cluster. 
This is the highest S/N observation of a soft excess cluster to date, 
and  we use for the first time these data to confirm 
the earlier detections and to determine the presence and characteristics
of the soft excess. In a companion paper (Mittaz, Lieu and Bonamente 2005) we use
the \xmm\/ data of this and other clusters to investigate the presence of OVII lines
associated with the soft excess.

The origin of the soft excess emission in galaxy clusters is yet uncertain.
There is incresing evidence in favor of a thermal origin of the soft excess, especially since the
report of OVII emission lines associated with the
soft emission of a few clusters observed by \xmm~(Kaastra et al. 2003).
A non-thermal origin of the phenomenon is also possible,
which requires a widespread presence of relativistic electrons
that scatter the cosmic microwave background
(Hwang et al. 1997; Sarazin and Lieu 1998; Lieu et al. 1999; Bonamente, Lieu and Mittaz 2001).

This paper is structured in the following way. In section \ref{data} we describe the \xmm~ data of \as,
with emphasis on the performance of the low-energy (E$<$2 keV) response
of the EPIC instruments. For this purpose we also analyze an \xmm~ observation of \pks, which we
use as a calibration target.
In section \ref{softexcess} we establish the presence of the soft excess emission in our \xmm~ data,
and provide the thermal and non-thermal modelling. 
In sections \ref{thermal} and \ref{nonthermal} we
discuss the physical interpretation of  the thermal and non-thermal
models, comparing our results with the large scale simulations
of Cen and Ostriker (1999) and of Cheng et al. (2004).
In section \ref{interpretation} we summarize our results, and indicate
that the thermal scenario is the favorite interpretation of these data. 
Finally, in section \ref{comparison} we compare our results with earlier works
on \as.

Throughout the paper we assume a Hubble constant of $H_0=70$ km s$^{-1}$ Mpc$^{-1}$,
and an $\Omega=1$, $\Omega_M=0.27$ and $\Omega_b$=0.05 cosmology.
At the
source's redshift of z=0.058, 1 arcmin corresponds to 72 kpc.
All quoted uncertainties are 90\% confidence, unless otherwise specified.

\section{XMM-Newton data of \as \label{data}}

\subsection{Data and data reduction }

\as~was observed by \xmm~on Nov. 20-21, 2002 for
122 ks, observation ID was 0147800101. 
The observation collected data with the EPIC imaging spectrometers (MOS1, MOS2 and pn)
featuring the thin filter.

The data was reduced using the SAS software provided by the \xmm~team, following the
standard procedure:
 the EPIC pipelines were run on the raw data, in order to apply the
most current calibration information, available through the CCF files
also provided by the \xmm~calibration team.
 For the MOS detectors (MOS1 and MOS2), events were selected having
{\it pattern}$<=$12 and {\it flag}=0. For pn, we selected events
with {\it pattern}=0 and {\it flag}=0. This is the same selection as
in Kaastra et al. (2003).
 
\subsection{Time-variable particle background \label{flares}}

The EPIC background comprises a variable component due to soft protons, which
is sometimes enhanced by orders of magnitude over the
quiescent rate. The energy band above $\sim$ 10 keV is dominated by this
radiation, and can therefore be used to monitor the presence of such particle background
`flares' (e.g., Arnaud et al. 2001; Lumb et al. 2002; Nevalainen et al. 2003).

We extract light curves of our \as\/ observation at energies $\geq$9.5 keV (MOS) and
$\geq$10 keV (pn), and discard those time intervals where the count rate
exceeds the quiescent count rate.
%, as determined by an analysis of several \xmm~
%observations (Nevalainen et al. 2005). 
The threshold levels for rejection
were set to respectively 108 and 112 counts (500 s)$^{-1}$ for MOS1 and MOS2, 
and 168 counts (500 s)$^{-1}$ for pn, similar to the levels of Kaastra et al. (2003).
This reduction process leaves respectively 69 ks, 67 ks and 64 ks of clean data
for MOS1, MOS2 and pn.

\subsection{Background subtraction \label{background}}

We use the background event files provided by the \xmm~Science Operation Center
for the purpose of background subtraction (Lumb et al. 2002). This is a collection of
8 blank-sky fields at high Galactic latitude, which we filtered with the same criteria
as the cluster field. In so doing,  we ensure that
the soft proton background (see section \ref{flares}) is correctly subtracted. 
The cosmic X-ray background, however, is the dominant background
component at the low energies (E$<$1 keV) where the
cluster soft excess radiation is present. This photonic component of the background
features sky variations  ($\sim 35$\% at $\sim0.25$ keV: 
Snowden et al. 1998; Kaastra et al. 2003), which need to be accounted for when
establishing the presence of the soft excess emission.

\begin{figure}
\includegraphics[scale=0.5,angle=-90]{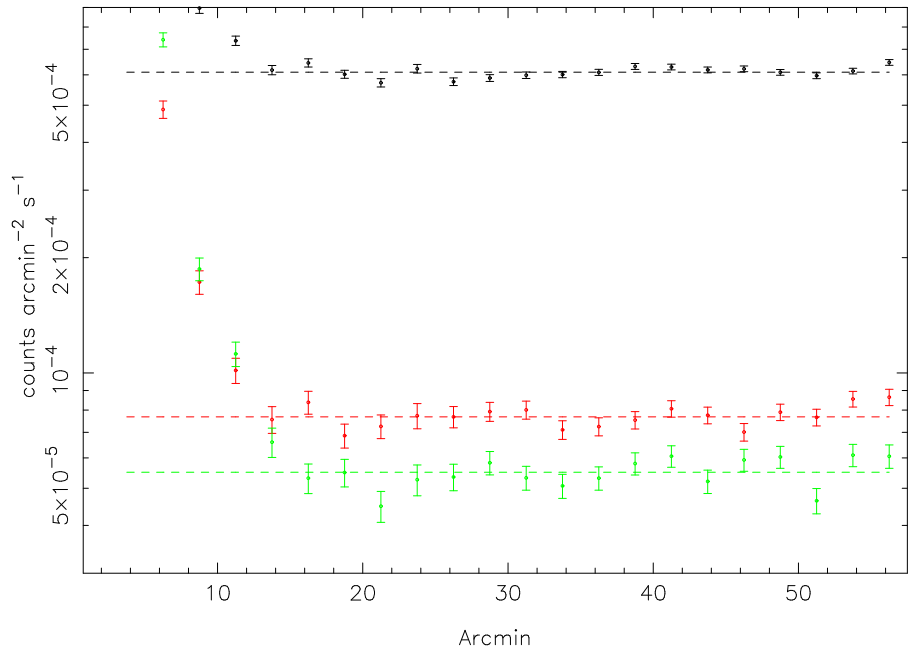}
\caption{Radial profiles of the \rosat\/ PSPC surface brightness.
In black, the R2 band emission; in red, the R4 band and in green the
R7 band. Dotted lines
are the best fit constant model to the 20-55 arcmin data.
\label{radial_pspc}}
\end{figure}

The \xmm\/ detectors cover an area of $\sim$14 arcmin radius ($\sim$ 1 Mpc) 
and therefore cannot
investigate the cosmic X-ray background in the neighborhood of \as. For this purpose,
we analyze a \rosat\/ pointed observation of the cluster, covering an area of $\sim$1 degree
radius ($\sim$ 4.3 Mpc). This is the same observation in which we initially discovered
the soft excess emission of \as\/ (Bonamente, Lieu and Mittaz 2001).
The radial profiles of the X-ray emission in the R2, R4 and R7 bands, respectively
$\sim$0.25 keV, $\sim$0.75 keV and $\sim$1.5 keV, 
are shown in Figure \ref{radial_pspc}. 
It clearly shows that the R2, R4 and R7 band surface brightness profile reaches a constant value at
$\sim$12 arcmin radius from the cluster's center.
The emission in all PSPC bands maintains a constant level of emission out to the limiting radius 
of the PSPC detector.

On larger scales,  
a low-resolution measurement of the local soft X-ray background is available through
the ROSAT All-Sky Survey maps (Snowden et al. 1998).
We obtain RASS images around \as~ and around the 8 pointings that constitute the \xmm~ background
through the NASA Skyview facility.
We use the R12 and R45 band images~\footnote{RASS bands R12 and R45 are similar
to bands R2 and R4 of the pointed ROSAT data.}
that cover the
$E\leq$1 keV band where the soft excess is strongest, and
observe that the 10 degree region around \as~
features a soft X-ray background $\sim$ 20\% higher than that of the Lumb et al. (2002) background fields
in both R12 and R45 band (Table  \ref{rass}), and no indication of a spatial gradient in the
R12 and R45 band emission. The 0-1 degree region is clearly affected by the diffuse
cluster emission, which is included in the RASS maps.
\begin{deluxetable}{lcccc}
\tablewidth{0pt}
\tablecaption{\rosat~ All-Sky Survey soft X-ray background data \label{rass}}
\tablehead{ Field      & 0-1 deg. & 1-2 deg. & 2-5 deg. & 5-10 deg.}
\startdata
\multicolumn{5}{c}{R12} \\
\as  & 1.09 & 0.99 & 0.99 & 1.02 \\
Background & 0.81 & 0.79 & 0.77 & 0.78 \\
Avg. back. & \multicolumn{4}{c}{0.79} \\
\hline
\multicolumn{5}{c}{R45} \\
\as & 0.18 & 0.14 & 0.13 & 0.13 \\
Background & 0.11 & 0.11 & 0.12 & 0.11 \\
Avg. back. & \multicolumn{4}{c}{0.11} \\
\enddata
\tablecomments{Fluxes are in units of $10^{-3}$ counts s$^{-1}$ arcmin$^{-2}$.
The resolution of the RASS maps are $\sim$40 arcmin, so that the central
1 degree region of the \as~ observation is affected by the hot ICM emission.}
\end{deluxetable}

The analysis of the ROSAT data around \as\ indicates that the cluster's
soft emission reaches the
background level at a distance of $\sim$12 arcmin from the cluster's center.
The available data cannot determine the nature of the $\sim$ 20\% enhancement of
the background near \as, with respect  to the background fields
of Lumb et al. (2002). 
For that purpose, spectroscopic observations 
with few eV resolution are needed
that measure the redshift of the
background emission. Such resolution will become available 
with ASTRO-E2.
The enhancement is likely of local (Galactic) origin, 
although large-scale emission at the cluster's redshift cannot be ruled out.

\begin{figure}
\includegraphics[scale=0.5,angle=-90]{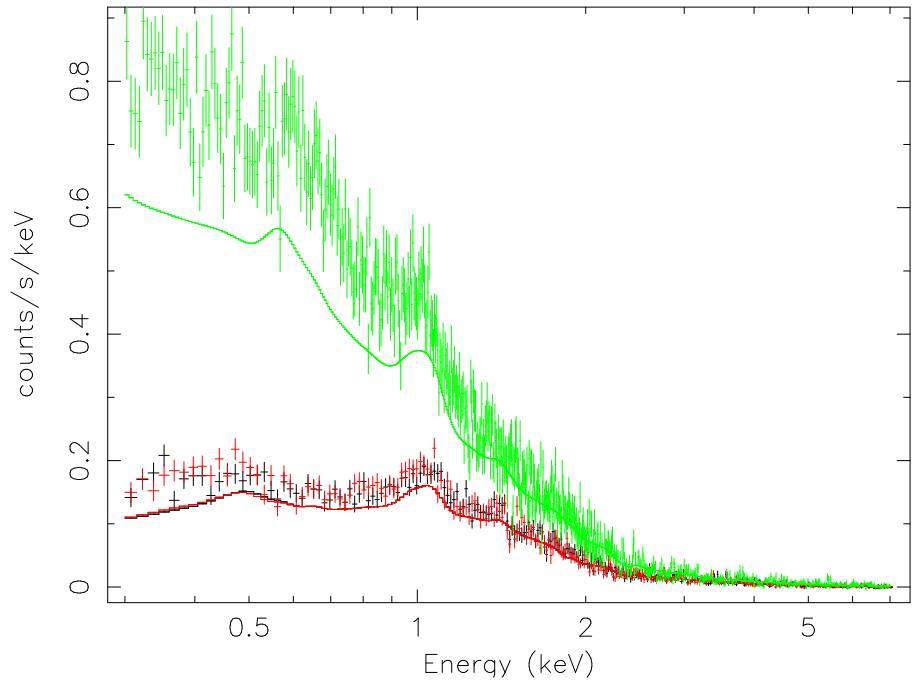}
\caption{\xmm~ spectra of the 4-6 arcmin region of
\as. Solid lines are the best-fit 2-7 keV 1-temperature models,
which were extrapolated to low energies. In black and red the MOS1
and MOS2 data, in green the pn data. The low-energy positive residuals
represent the soft excess emission.
\label{as1101_4-6}}
\end{figure}

In this paper we follow a conservative approach, and assume that the enhancement in the background
is entirely of local origin. We therefore use the 12-14 arcmin \xmm\/ 
data to model this background enhancement,
and subtract this additional background component from the \as\/ spectra. This method
is equivalent to that of Mittaz, Lieu and Bonamente (2005), in which we use \xmm\/ data to investigate
in more detail the presence of OVII and OVIII emission 
lines in the soft excess spectra of a few galaxy clusters, including \as.
Kaastra et al. (2003) followed a different approach, and opted to use the background fields
of Lumb et al. (2002) with the addition of a systematic error in an effort to account
for possible variations of the soft X-ray background.
Our choice of using the 12-14 arcmin
region to estimate the local soft X-ray background
results in no detection of OVII emission lines from \as.
This conclusion  is evident in the 4-6 arcmin spectrum of
\as\ (Figure \ref{as1101_4-6}), which shows
no evidence of an emission feature at $E\sim 0.5$ keV. Similar results
apply to the 6-9 and 9-12 arcmin annuli.
Further details on the presence of emission lines associated with the
soft excess can be found in Mittaz, Lieu and Bonamente (2005).

\subsection{Assessment of EPIC's soft X-ray calibration \label{calibration}}

We limit our analysis to the 0.3-7 keV energy range, where the accuracy of
the EPIC effective area calibration is $\leq$10 \% (Kirsch 2004).
We assess directly the soft calibration of the EPIC detectors through the analysis
of the 
\xmm~ observation of \pks, a bright BL Lac source observed with the thin filter 
for 58.5 ks (observation ID 0080940301).
The observation, reduced following the same steps as the cluster observation,
features respectively 24 ks  and 33 ks of useful  MOS2 and pn data (MOS1 was used in 
timing mode). This observation was affected by event pile-up near the detector
aimpoint position. In order to eliminate this unwanted effect, our spectra were
extracted from an annular region that excludes the aimpoint. This choice renders 
our data free of pile-up events.

BL Lac objects are variable, and their X-ray spectra are generally well fit by
power-law models (e.g., Ravasio et al. 2002; Wolter et al. 1998).
A spectral break is sometimes needed to account for spectral curvature (e.g., Perri
et al. 2003; Chiappetti et al. 1999).

\begin{figure}
\includegraphics[scale=0.5,angle=-90]{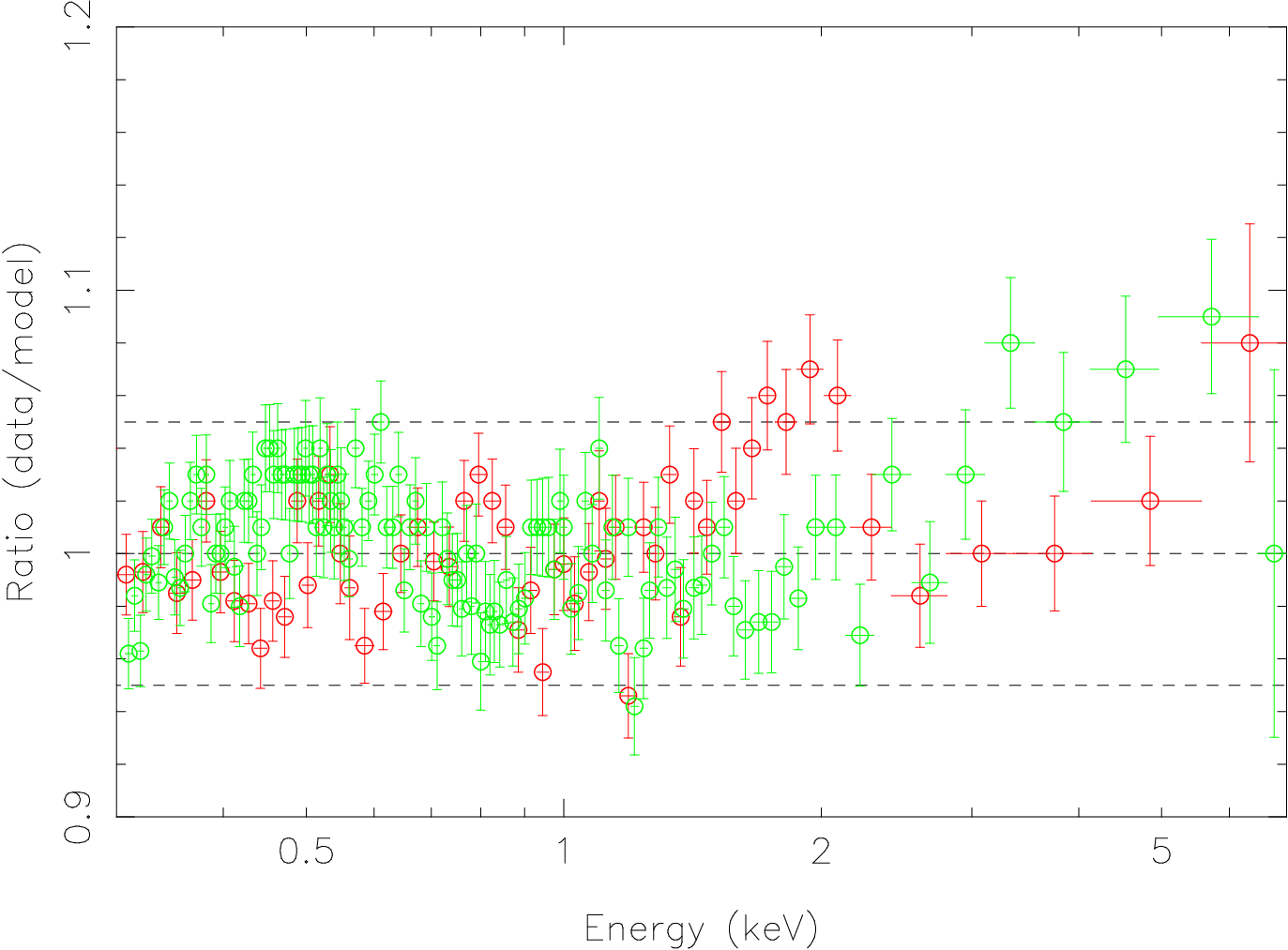}
\includegraphics[scale=0.5,angle=-90]{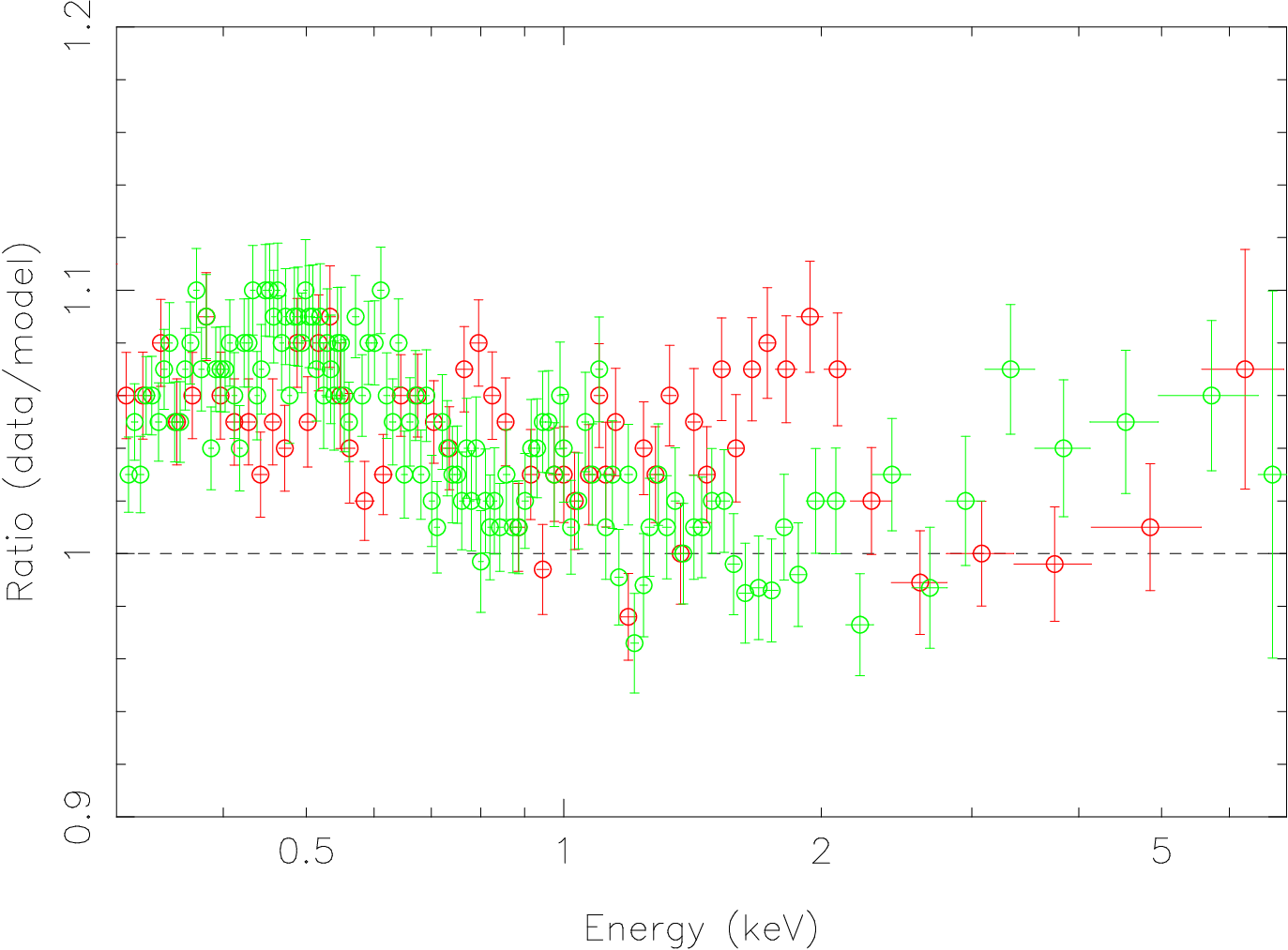}
\caption{
(a) Residuals of the 0.3-7 keV power-law model
and  (b) of the 2-7 keV power-law model of  \pks.
In green is the pn data and in red the MOS2 data.
For clarity, after the fit the residuals were rebinned to achieve a S/N=50.
 \label{fig_pks}}
\end{figure}

We fit the \pks~data in the same way in which we will subsequently fit the cluster data.
We fit jointly the MOS and pn spectra, after ascertaining that the individual spectra
are consistent with one another. Spectra were binned with a minimum of 25 counts per
channel.
First, we fit the full 0.3-7 keV range with an absorbed power-law model,
fixing the $N_H$ column density at the Galactic value of 1.36$\pm0.10 \times 10^{20}$ cm$^{-2}$
(Dickey and Lockman 1990).
The model fits the data well, and the low energy residuals are $\leq$ 5\% (Fig. \ref{fig_pks}(a)).
We then fit the same model to the 2-7 keV data.
In Table \ref{tablepks} we show the results of the 
fits: the two fitting methods yield consistent results,  indicating that the source
follows a simple power law in the selected energy range, as already known from a previous
study (Kaastra et al. 2003).~\footnote{A broken power law model provides no improvement
to the 0.7-7 keV fit of our spectrum, providing further indication that the spectrum is
well described by a single power law. Even in those epochs when the source could
not be fit by a simple power law model, e.g., in the BeppoSAX observation of
Giommi et al. (1998), the residuals from the best-fit single power law model
did not exceed 20\% in the entire 0.7-7 keV band, and were $<$10\% in the softest
X-ray channels (0.3-1 keV).}
\begin{deluxetable}{cccc}
\tablewidth{0pt}
\tablecaption{\xmm~observation of the calibration source \pks \label{tablepks}}
\tablehead{$\alpha$ & Norm. MOS2 & Norm. pn & $\chi^2_r$(dof) }
\startdata
\multicolumn{4}{c}{0.3-7 keV fit} \\
2.793$\pm$0.004 & 0.0341$\pm$0.0001 & 0.0248$\pm$0.0001 & 1.23(1085) \\
\hline
\multicolumn{4}{c}{2-7 keV fit} \\
2.760$\pm$0.030 & 0.033$\pm$0.001 & 0.024$\pm$0.001 & 1.03(626) \\
\hline
\enddata
\begin{center}
\footnotesize{Note: The normalization constant of the power-law model is in the XSPEC
customary units.}
\end{center}
\end{deluxetable}
We now extrapolate the best-fit 2-7 keV model down to 0.3 keV, in order to investigate the
low-energy residuals. The average residuals in the 0.3-1 keV band are respectively +5\% and
+6\% for MOS2 and pn, as shown in Figure \ref{fig_pks}(b), with deviations
of less than $\sim$10\%. We therefore conclude that the level
of systematic uncertainties of the EPIC detectors 
at $E<1$ keV is less than 10\%, and that cluster soft excess fluxes 
in this band are real when they exceed that figure. A steep 
power-law model like that of \pks\/ exaggerates any positive low energy
residuals relative to a thermal cluster spectrum, thereby providing a conservative
test of the low-energy calibration.
A difference by a few \% in the 0.3-1 keV soft excess fluxes of MOS and pn is
also within the calibration uncertainties.

Alternatively, we fit the 0.3-7 keV spectra with a power-law model of variable 
absorbing column. The resulting  power-law index is consistent with the
numbers of Table \ref{tablepks}, and the best-fit column density is consistent with
the Galactic value. 
This result indicates that the apparent deficit
in the absorbing column seen in this cluster
(section \ref{simplemodels})
and in others (Kaastra et al. 2003) 
is  not an artifact due to calibration uncertainties,
but is the results of a genuine excess of soft photons.

\section{Presence of soft X-ray excess emission in the 	EPIC data of~\as \label{softexcess}}

\subsection{Spectral extraction and fitting}

We extracted spectra in 8 annuli of outer radii 1, 2, 3, 4, 6, 9, 12 and 14 arcmin.
We use the outermost annulus (12-14 arcmin) only for the purpose of obtaining a model
of the local soft X-ray background, as described in section \ref{background}.
Spectra were extracted in 15 eV bin, and then
 rebinned to ensure that at least 25 counts were present in each bin.
We used the XSPEC software for the spectral analysis, the optically thin thermal model
was MEKAL, and the Galactic absorption cross sections those of Morrison 
and McCammon (1983)~\footnote{
See Bonamente, Lieu and Mittaz (2001) for discussion on the available cross section compilations.}.
The absorption column was fixed at the Galactic value ($1.8\times10^{20}$ cm$^{-2}$; Dickey
and Lockman 1990),
except in one case explicitly discussed in section \ref{simplemodels}.
We fit jointly the MOS1, MOS2 and pn spectra, enforcing the same normalization constants
between the MOS1 and MOS2 spectra, and allowing the pn spectrum to have  a separate
normalization. We use the 0.3-7 keV energy range in the spectral fitting.

\subsection{Simple models of the \as~ spectra \label{simplemodels}}
\subsubsection{1-temperature model with central `cooling' gas\label{1t}}
\begin{figure}
\includegraphics[scale=0.35,angle=-90]{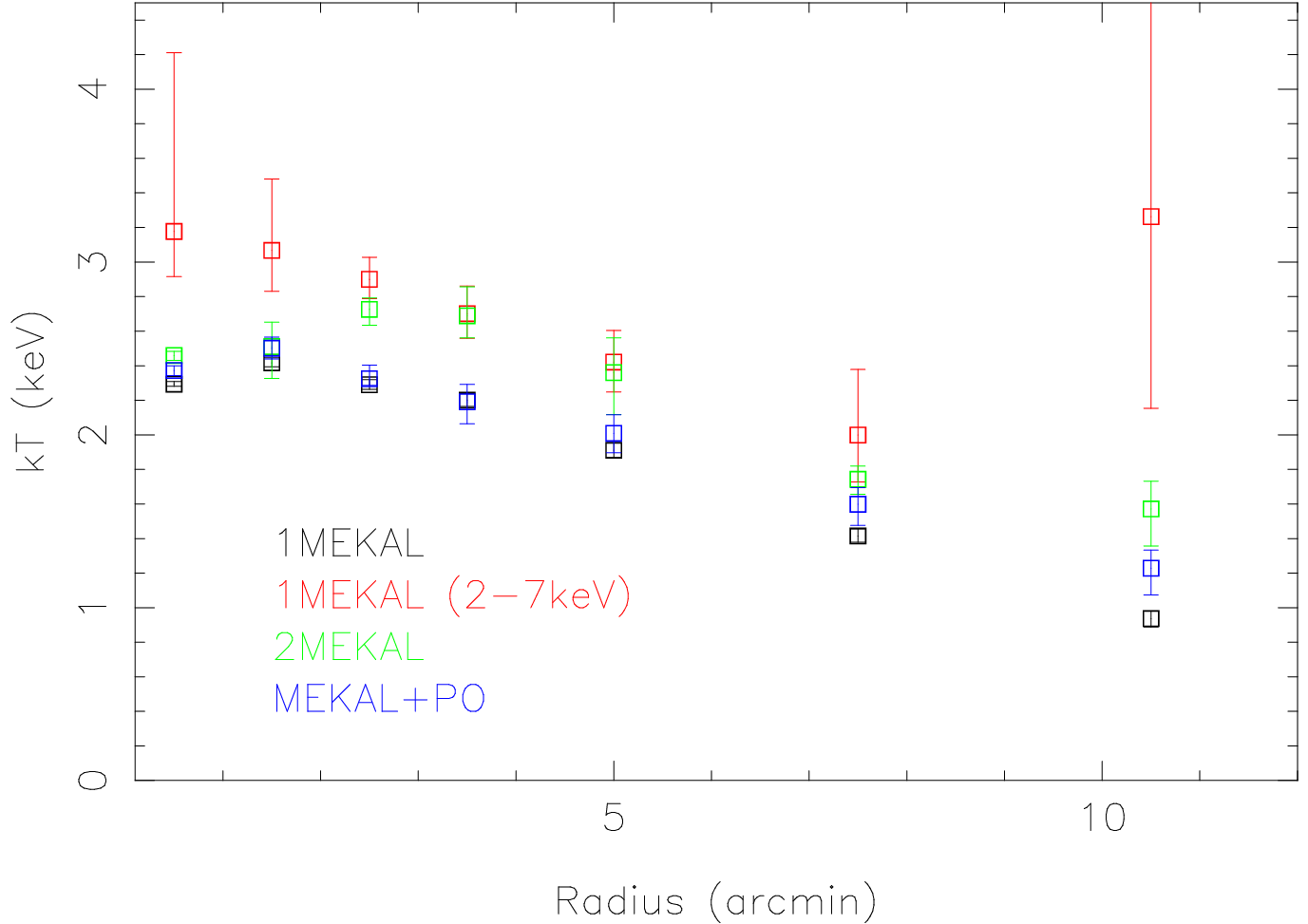}
\includegraphics[scale=0.35,angle=-90]{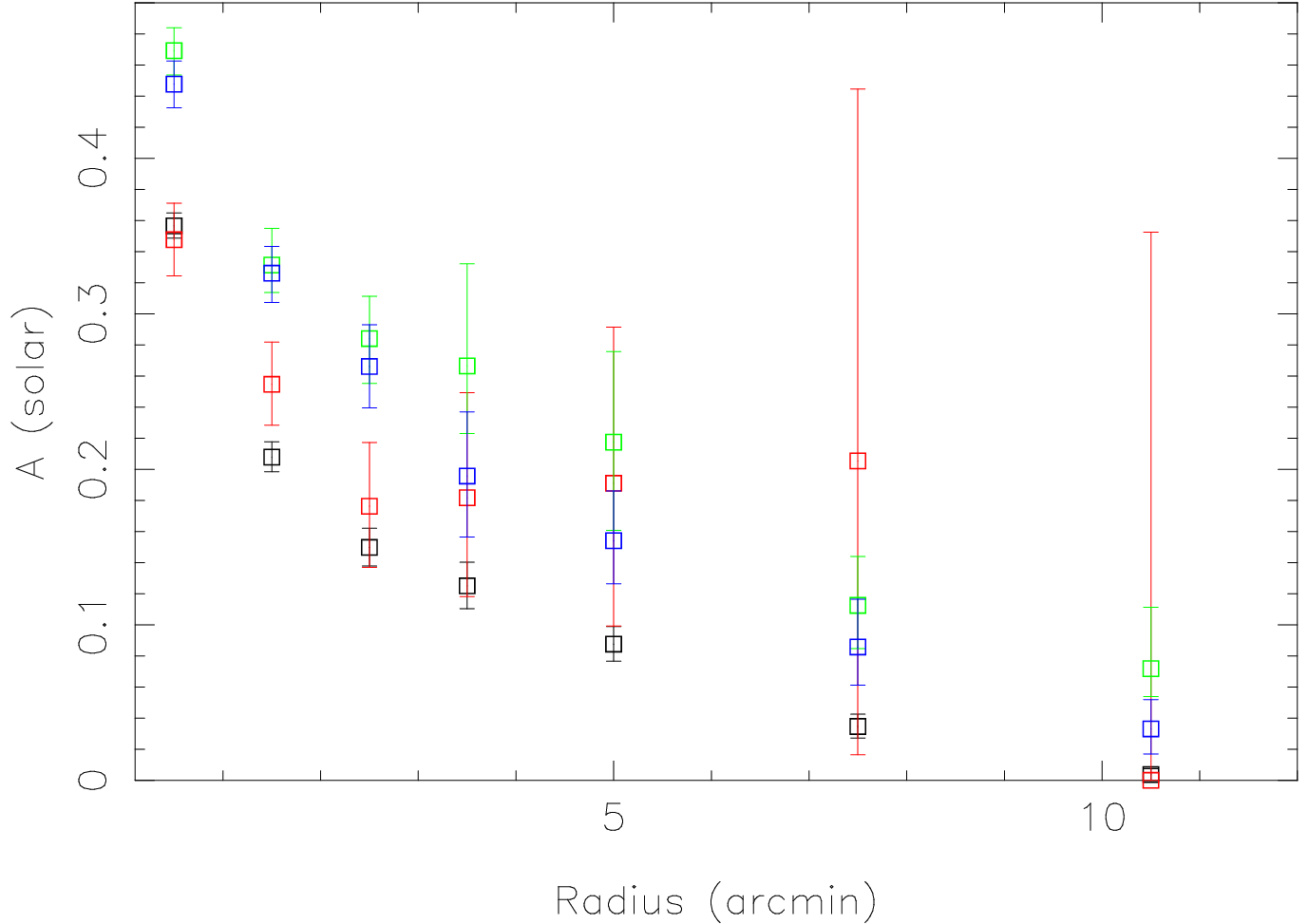}
\includegraphics[scale=0.35,angle=-90]{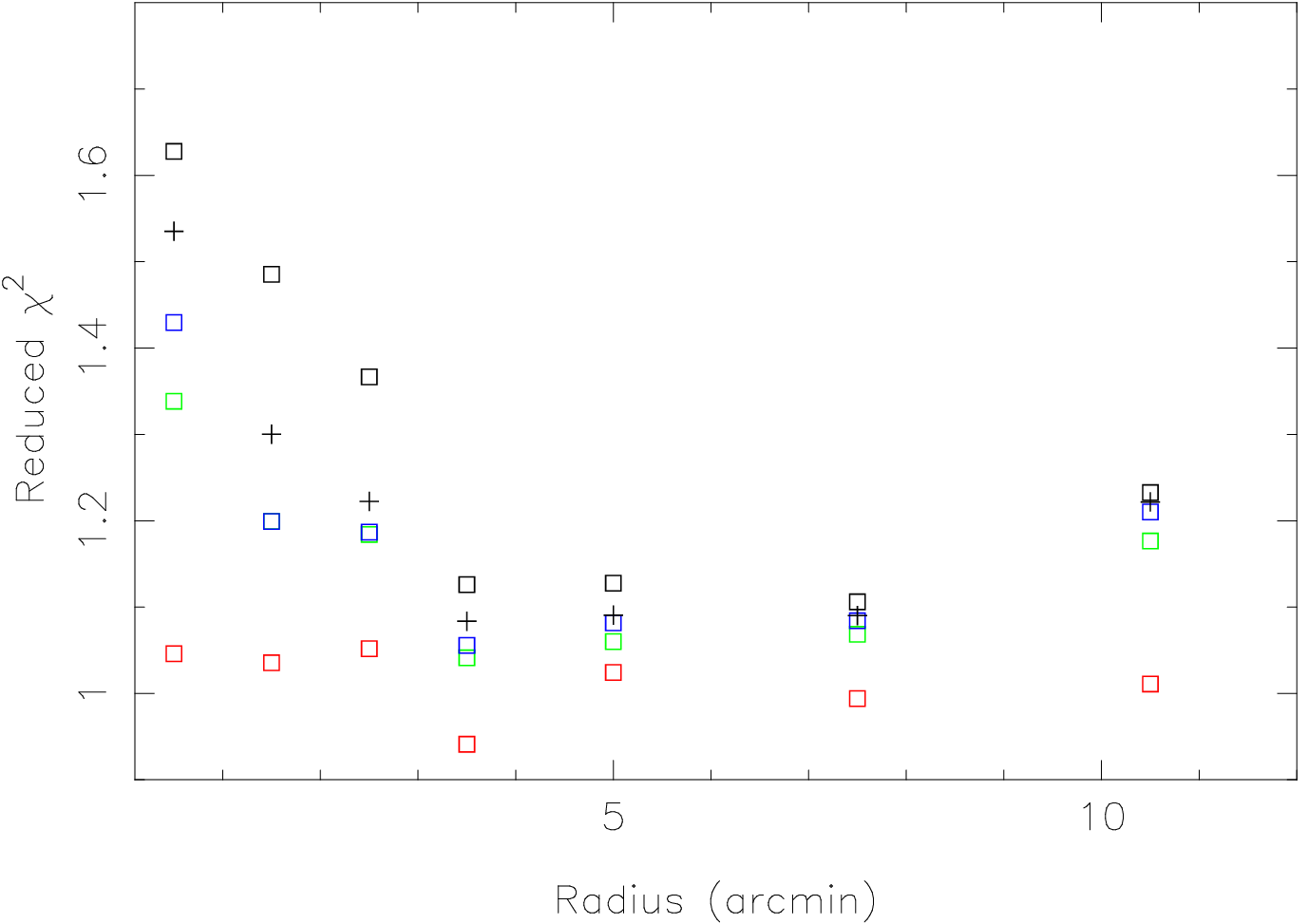}
\caption{
Temperature, abundance
and reduced $\chi^2$ profiles for the 4 models
considered in sections \ref{simplemodels} and \ref{complexmodels}.
In the $\chi^2$ panel we also added the result of the
variable-$N_H$ model (crosses).
\label{profiles}}
\end{figure}

\begin{figure}
\includegraphics[scale=0.7,angle=-90]{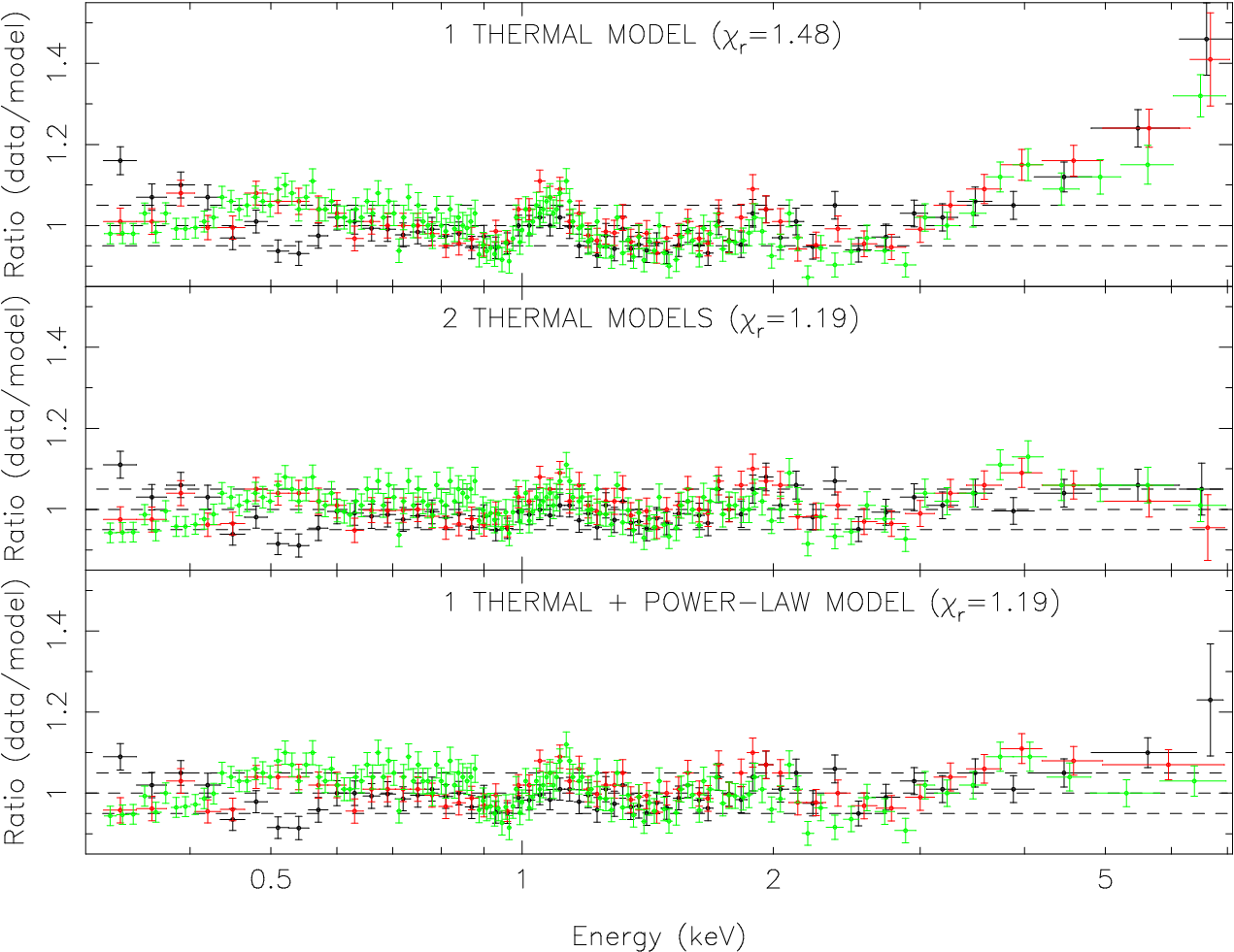}
\caption{Residuals of the (a) 1-temperature model of section \ref{1t},
(b) 2-temperature model of section \ref{2t} and (c) non-thermal
model of section \ref{1tpo} for the 1-2 arcmin annulus.
In black and red the MOS1 and MOS2 data, and in green the pn data.
For clarity, after the fit the residuals were rebinned to achieve a S/N=30.
\label{A}}
\end{figure}

\begin{figure}
\includegraphics[scale=0.7,angle=-90]{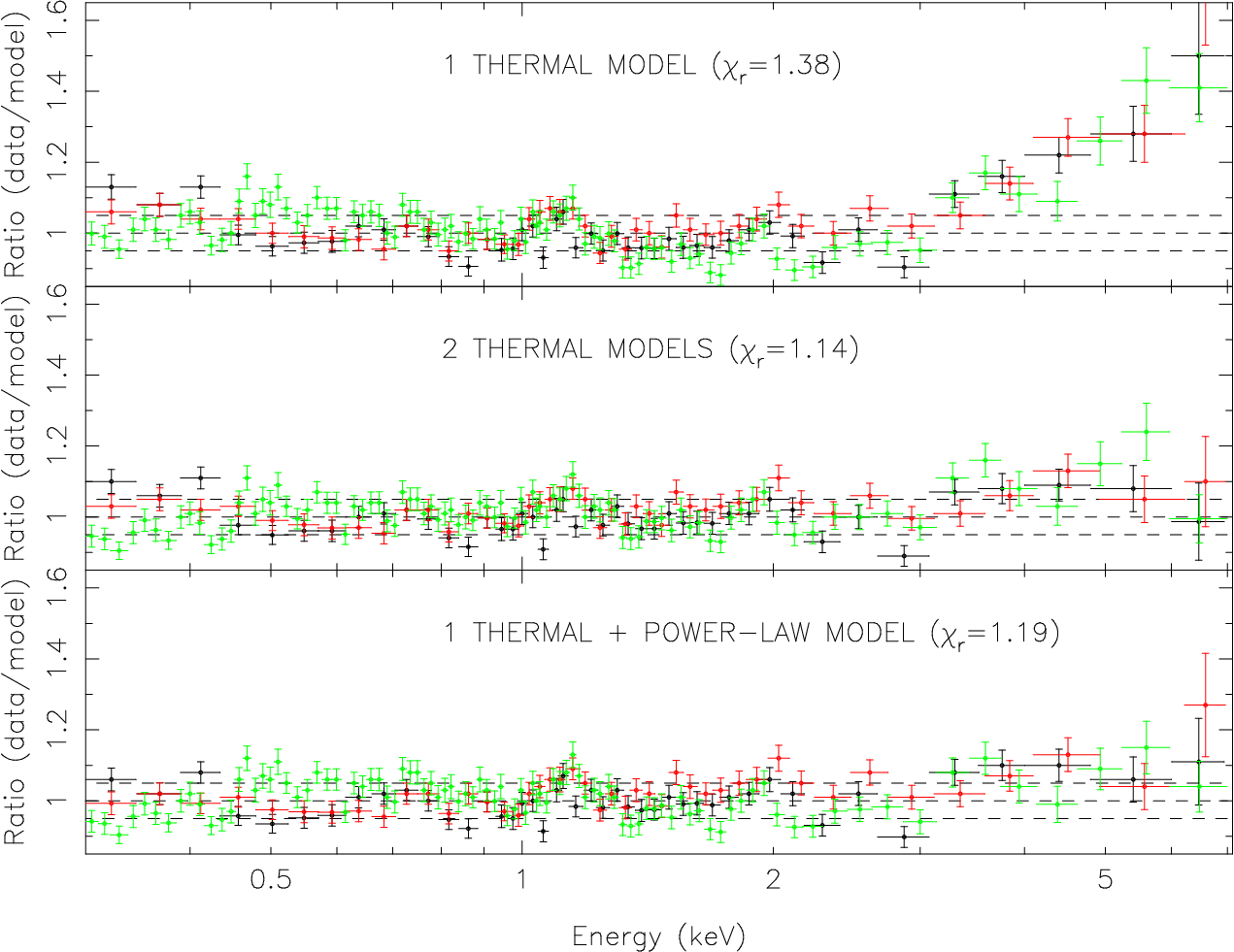}
\caption{Residuals of the (a) 1-temperature model of section \ref{1t}, (b) 2-temperature model of section \ref{2t} and (c) non-thermal
model of section \ref{1tpo} for the 2-4 arcmin annulus.
In black and red the MOS1 and MOS2 data, and in green the pn data.
For clarity, after the fit the residuals were rebinned to achieve a S/N=30.
\label{B}}
\end{figure}

First, we attempt to fit the 0.3-7 keV spectra with a simple thermal model.
The innermost regions of \as~ host cooler gas, due to the cooling of the
hot ICM. 
Earlier \xmm~ observations
(Kaastra et al. 2001; Kaastra et al. 2004a; Peterson et al. 2003) revealed
that the cooler gas in \as~ is confined to the central 2 arcmin.
We account for this cooler gas as in Kaastra et al. (2003), i.e., by adding a second
phase at 1/2 the temperature of the hot ICM in the innermost two annuli (0-1 and 1-2 arcmin
radius).
This simple model fails to describe the data accurately and
 results in very steep temperature and abundance gradients (see Figure \ref{profiles}).
A steep temperature profile was also found in an earlier \xmm~ observation of
\as~ by Kaastra et al. (2001), where the spectra were fitted to a similar
1-temperature model.
In Figure \ref{A}(a) and \ref{B}(a) we show the residuals from the best-fit
1-temperature model for two representative annuli. The residuals exceed the 
$\pm$5\% level seen in the case of the calibration target \pks~(Figure \ref{fig_pks})
and follow large-scale trends such as a negative deviations between 1.2 and 3 keV and
positive deviations at low and high energies, especially evident in the higher quality
pn data (green in Figures  \ref{A} and \ref{B}).
This behavior indicates the need for an additional spectral component, as
will be discussed in section \ref{complexmodels}.

Alternatively, we tried to account for cooler gas in the central regions using a classical
cooling flow model, and fitted the 0.3-7 keV spectra of the inner annuli with a 
1-temperature thermal model plus a cooling flow model.
The fits are not acceptable and of same goodness-of-fit as those obtained
using the 1-temperature thermal model plus the additional phase at 1/2 the temperature
described above in this section.
\subsubsection{1-temperature model with central `cooling' gas and with variable $N_H$}
\begin{figure}
\includegraphics[scale=0.5,angle=-90]{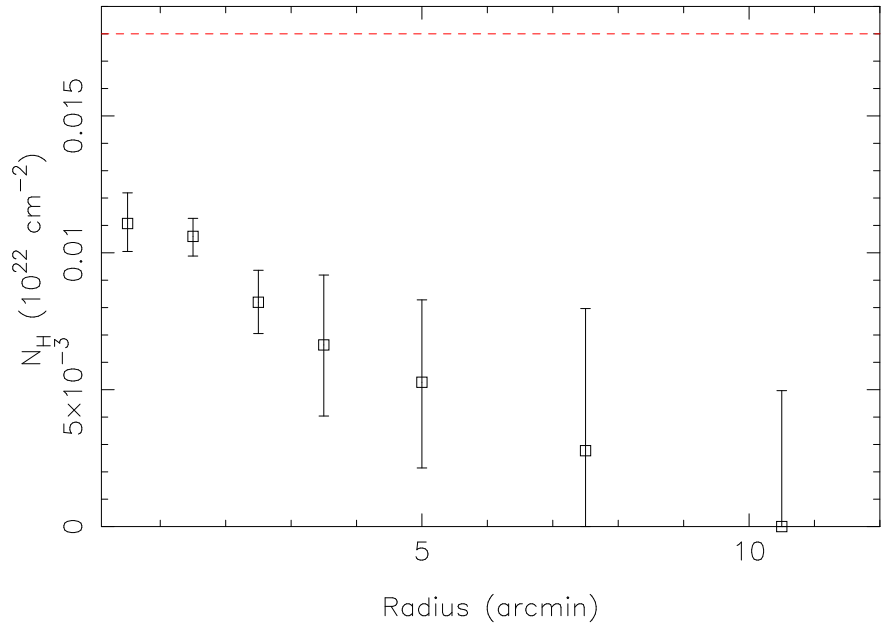}
\caption{
Best-fit HI column density of \as.
The Galactic column density is 1.8$\times 10^{20}$ cm$^{-2}$ (dotted line).
\label{nhprofile}}
\end{figure}

We investigate the possibility of modelling the \as~ spectra with the same 
1-temperature thermal model of section \ref{1t},
with the absorption column density as a free parameter. Although the best-fit $\chi^2$ statistic
improves, the fit is physically unacceptable, as the best-fit 
absorption column is significantly {\it sub}-Galactic
(Figure \ref{nhprofile}). The best-fit $N_H$ is  consistent with null Galactic absorption in
certain annuli.
A similar result was found in the data of the Coma cluster
(Bonamente, Joy and Lieu 2003) and of several other clusters observed by Kaastra et al. (2003), 
and is indicative of an excess of soft X-ray radiation with respect to this simple 1-temperature
hot ICM model. 
\subsubsection{1-temperature model fitted to the 2-7 keV band \label{2-7fit}}
\begin{figure}
\includegraphics[scale=0.5,angle=-90]{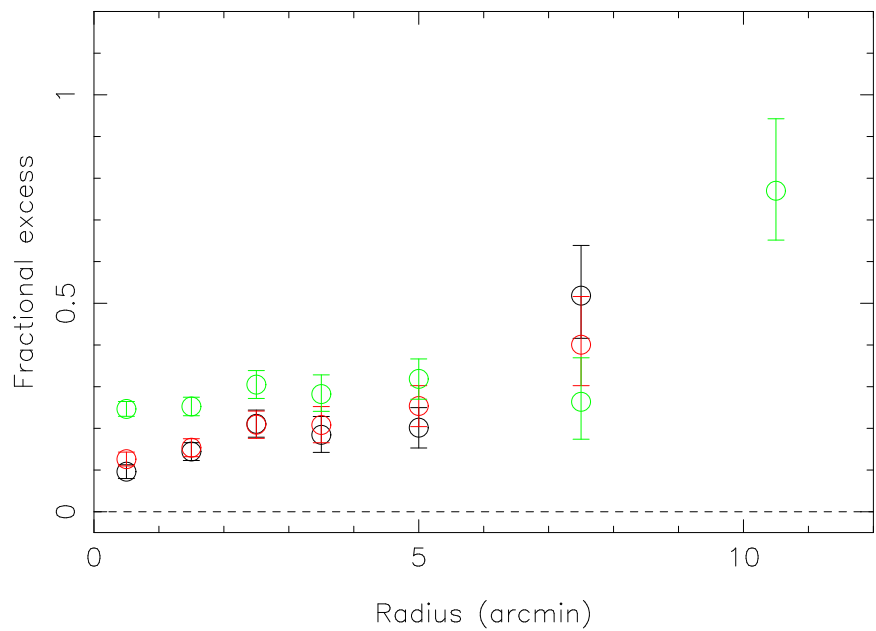}
\caption{Average fractional excess $\eta=\frac{O-P}{P}$ in the 0.3-1 keV band,
where $O$ is the observed soft flux, and $P$ is the
prediction from the hot ICM model.  The best-fit model
was determined by a 2-7 keV fit to the spectrum.
Statistical errors in the best-fit
hot ICM model and in the Galactic $N_H$ were properly accounted.
MOS1, MOS2 and pn data are respectively in black, red and green.
The 9-12 annuls has the MOS data omitted, as the 2-7 keV MOS data
do not properly constrain the hot ICM model.
\label{fracxs}}
\end{figure}

\begin{figure}
\includegraphics[scale=0.35,angle=-90]{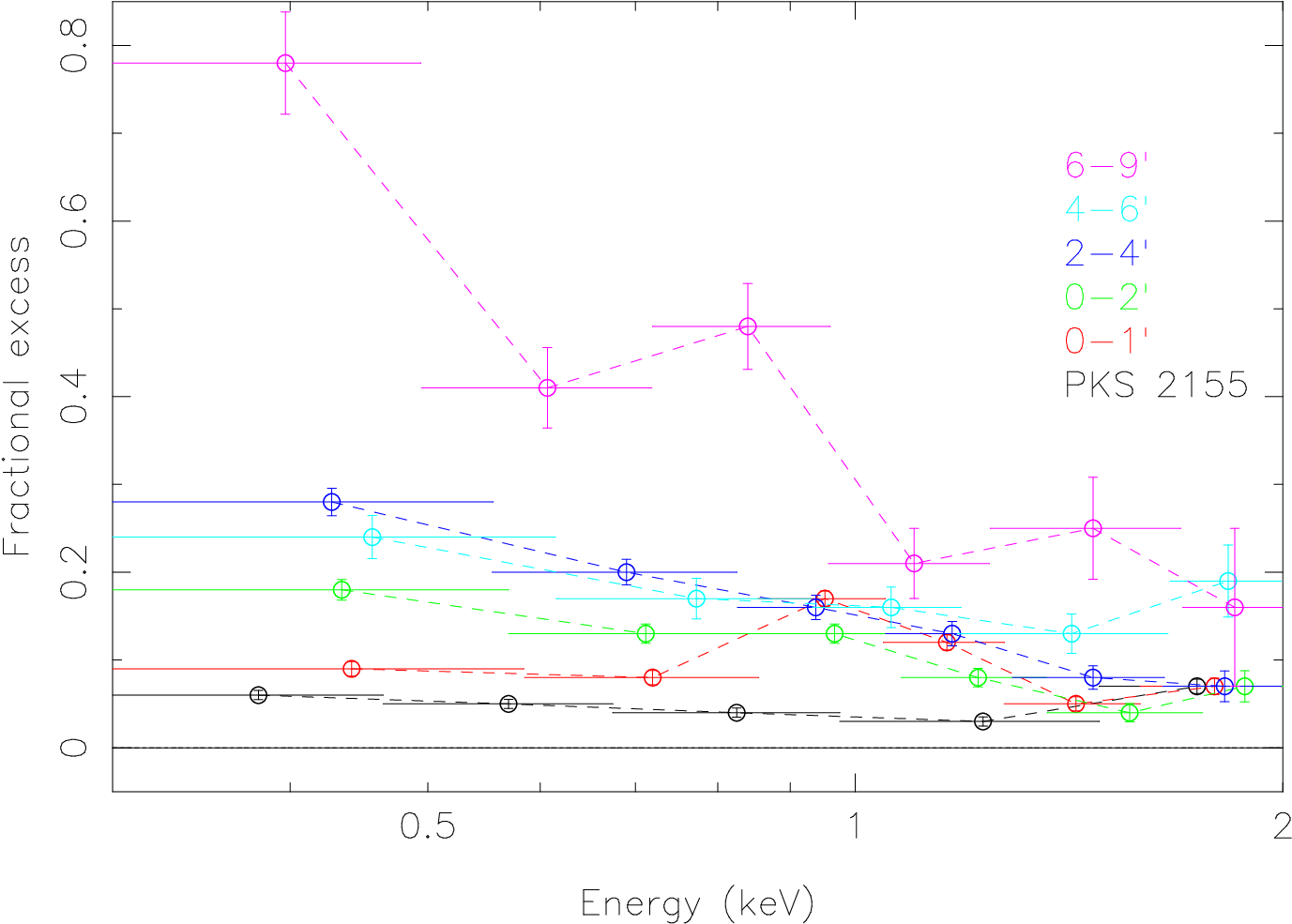}
\includegraphics[scale=0.35,angle=-90]{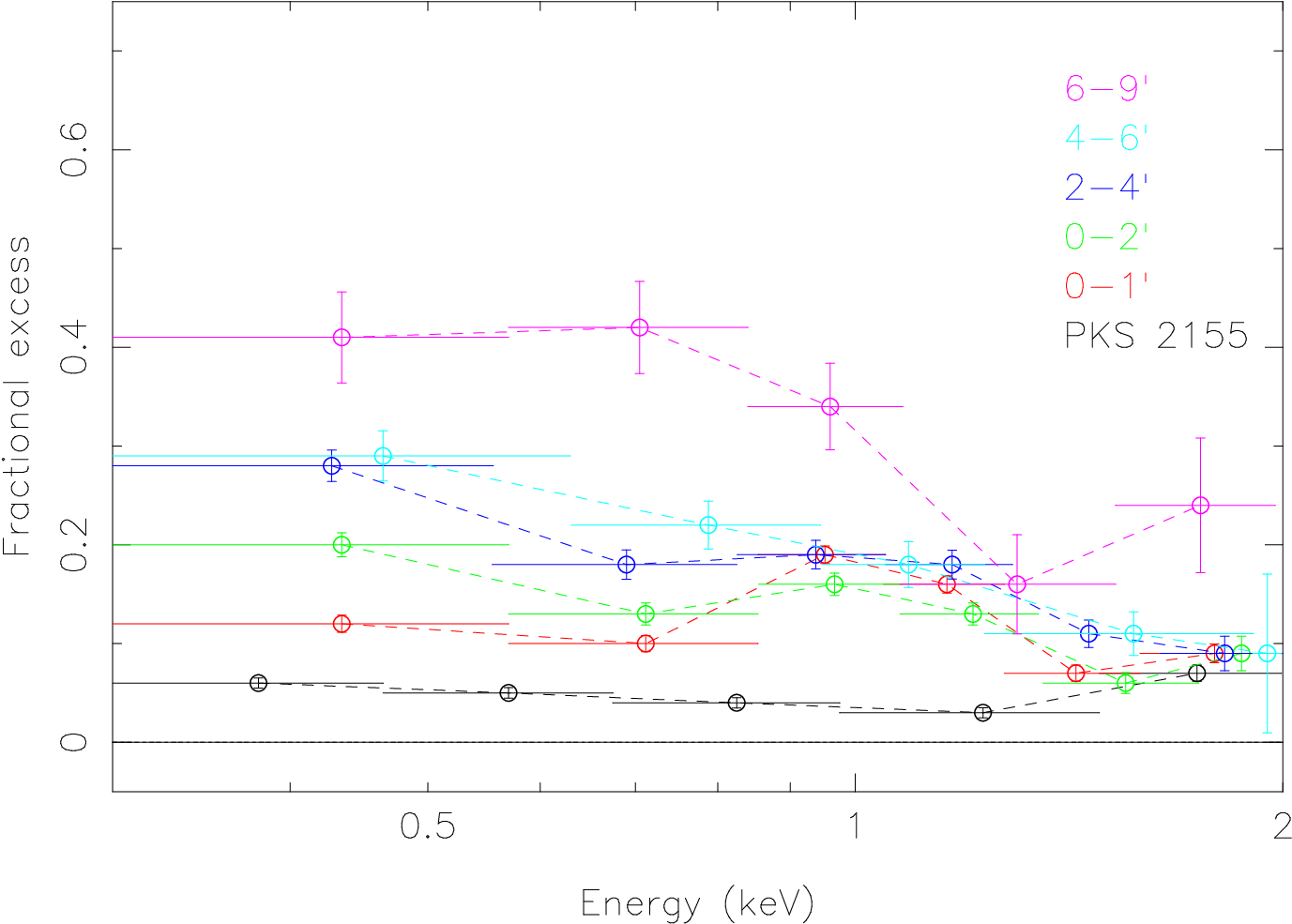}
\includegraphics[scale=0.35,angle=-90]{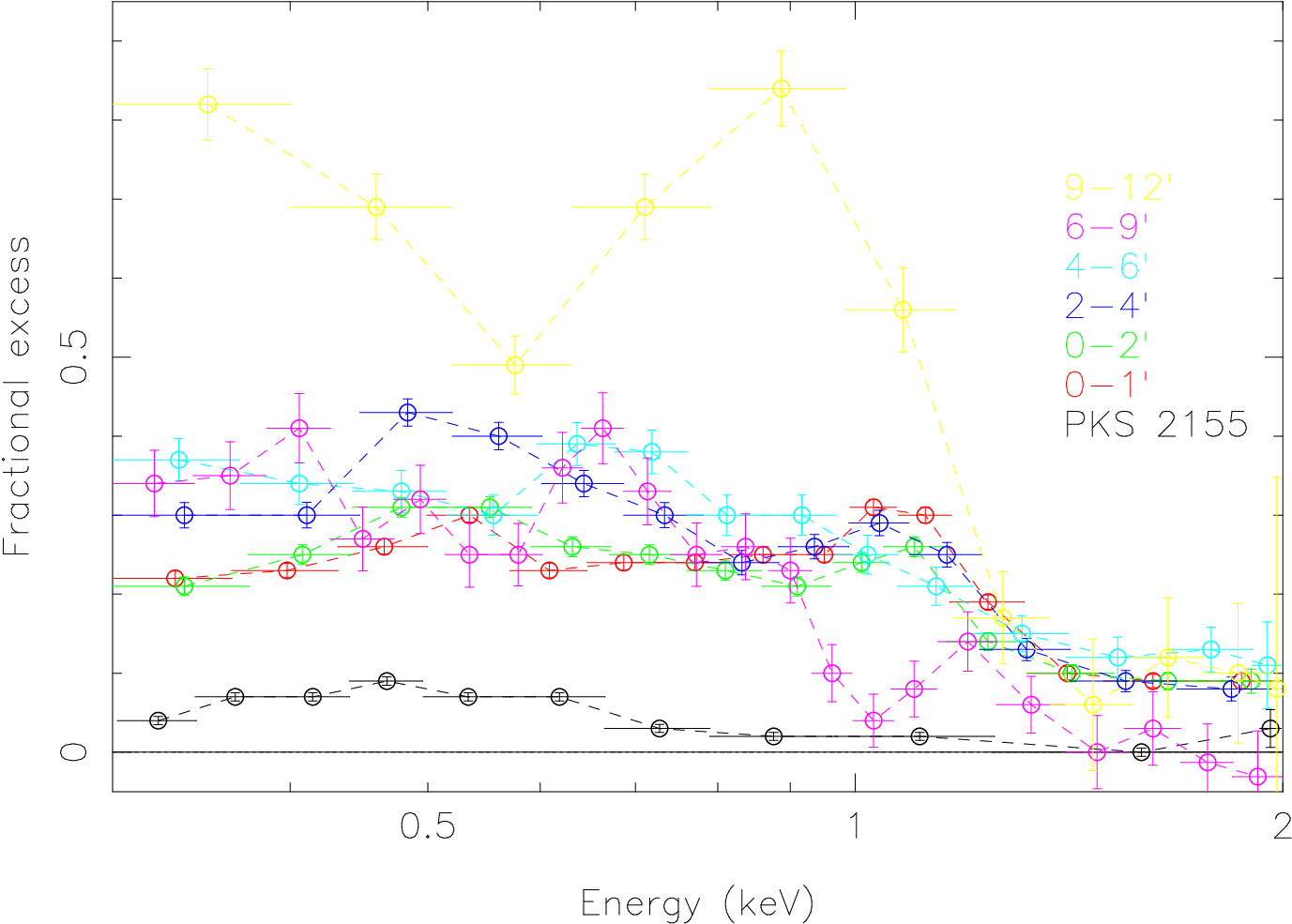}
\caption{Fractional excess $\eta$ in (a) MOS1, (b) MOS2 and (c) pn
for all annuli. The best-fit model
was determined by a 2-7 keV fit to the spectrum.
For clarity, after the fit the residuals were rebinned to achieve a S/N=30.
In black, we report the apparent excess of the calibration source \pks\/
(from Figure \ref{fig_pks}), which sets the level of the calibration
uncertainty. The error bars are the statistical errors in the
detected counts only.
\label{C}}
\end{figure}

Next, we discard the low-energy portion of the spectra (E$<$2 keV) from the fit, 
and fit the 2-7 keV data to the simple 1-temperature thermal model. In this case, we are
no longer sensitive to the presence of the central `cooling' gas, and we drop the
additional thermal model (at 1/2 the temperature of the hot plasma) from the 0-1 and 1-2 arcmin
annuli.
The results are now statistically acceptable, and the radial profiles of
temperature and abundance are less steep than in the case of the 0.3-7 keV fit above
(Figure \ref{profiles}).
In order to assess the presence of the soft excess emission,
we extrapolate this best-fit hot ICM model at lower energies
in the same way as we did for the calibration source \pks. 
Strong soft excess emission is found in all
annuli, as shown in Figure \ref{as1101_4-6} for the 4-6 arcmin region.
In Figures \ref{fracxs} and \ref{C} we show the fractional excess in the 0.3-1 keV band,
defined as $\eta=\frac{O-P}{P}$, where $O$ is the observed soft flux, and $P$ is the
prediction from the hot ICM model. 
The soft excess exceeds the level of the calibration uncertainties
(see section \ref{calibration})
and it is therefore due to a genuine excess of soft X-ray radiation with respect
to the 2-7 keV thermal model.
It is important to stress that this model is not used to measure the soft excess
flux, but only to establish the presence of an excess emission at low
energy that cannot be attributed to calibration uncertainties.
Soft excess fluxes will be measured by the models of section \ref{complexmodels}.
In fitting the 2-7 keV band alone we have ignored the Fe-L complex, which is an important
temperature diagnostic of the X-ray spectra. This model is also presented for the
sake of comparison with earlier results, e.g., those of Nevalainen et al. (2003), where
the authors fit the 2-7 keV \xmm\/ spectra of A3112, A1795 and Coma and find
a fractional excess  $\eta \simeq 10-40$ \% (see Figures 4 and 5 of
Nevalainen et al. 2003). 

\subsection{More complex models of the \as~ spectra \label{complexmodels}}

Having established the presence of an extra soft component in the \xmm~ spectra
of \as, we employ an additional component in the spectral fitting in order to
account for the excess emission.
\subsubsection{2-temperature model with central `cooling' gas \label{2t}}
\begin{figure}
\includegraphics[scale=0.35,angle=-90]{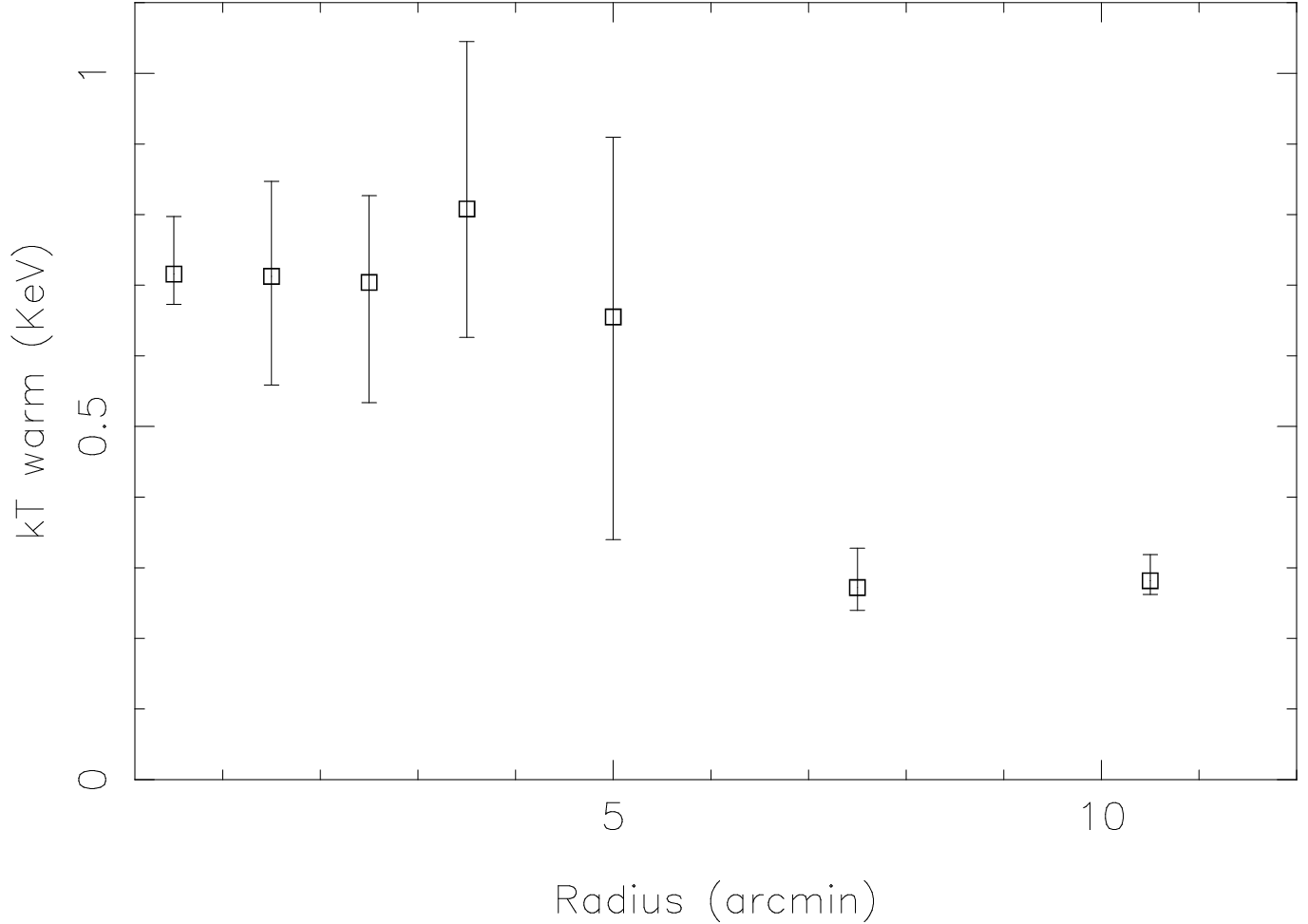}
\includegraphics[scale=0.35,angle=-90]{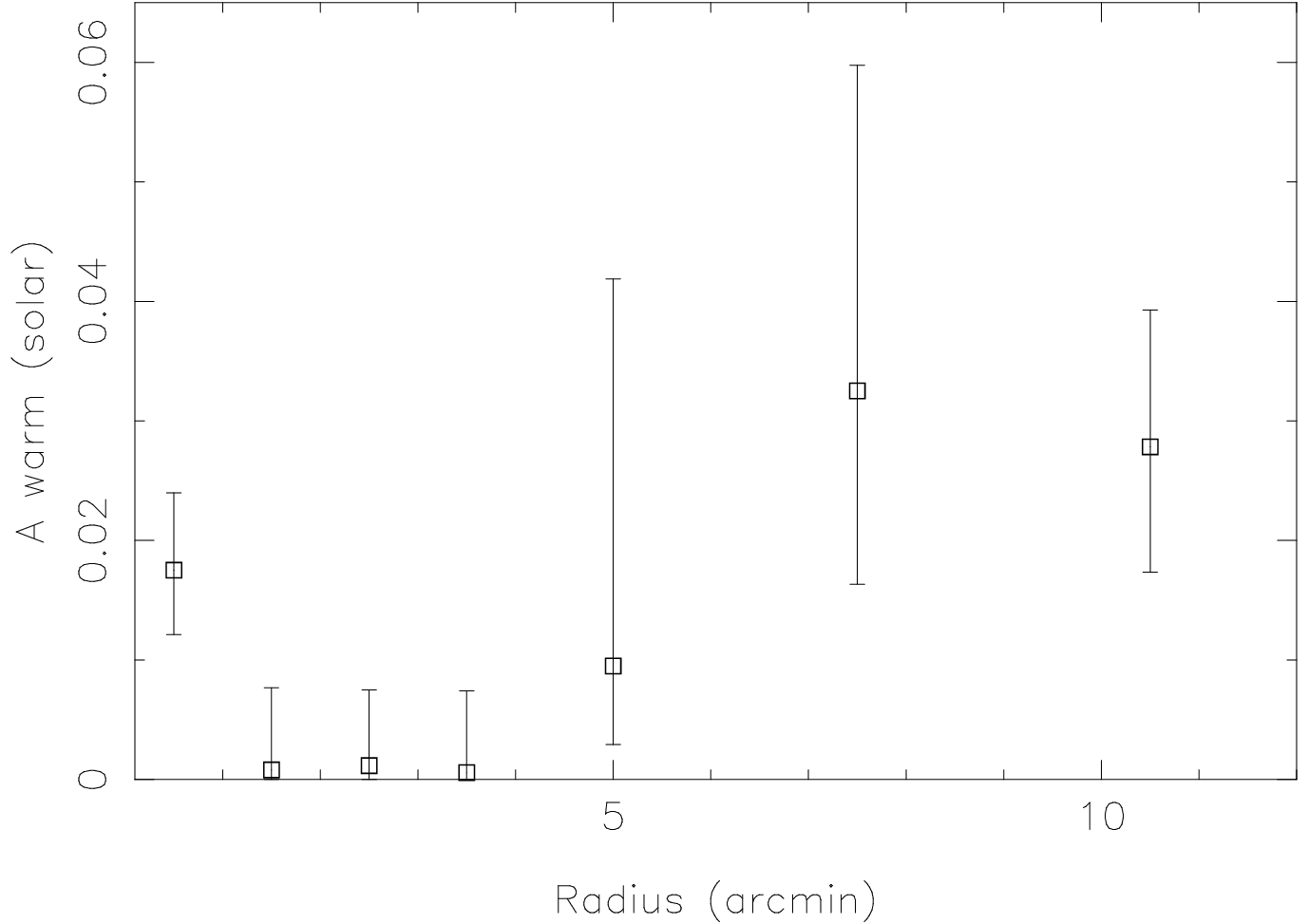}
\caption{
Radial profiles of the temperature and abundance
of the warm component from the model of section \ref{2t}.
%For ease of comparison, we also plot the
%normalization of the hot component as from Figure \ref{profiles}.
%In panels 3 and 4, black is  MOS data and red is PN data.
The warm model of annuli 6-9 and 9-12 had the same
normalization
constant for MOS and pn, given the limited number of counts in
these outermost annuli.
\label{warmprofiles}}
\end{figure}

\begin{figure}
\includegraphics[scale=0.4,angle=-90]{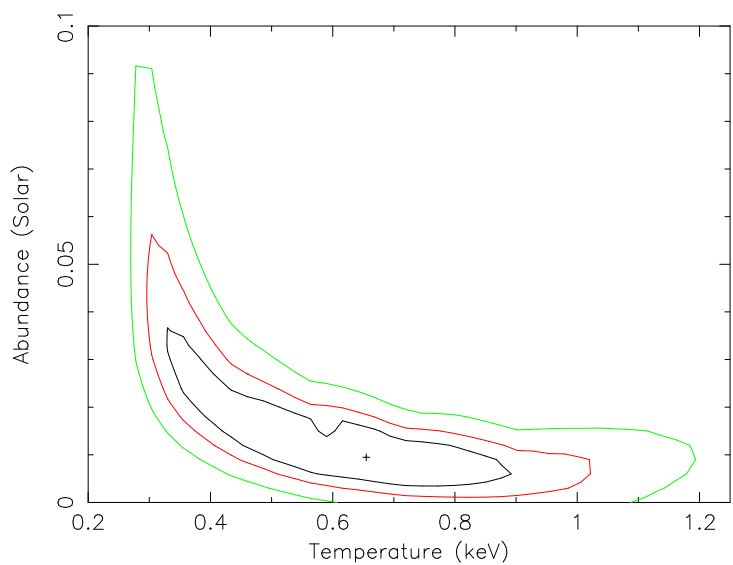}
\caption{Countour plot of the warm gas abundance and temperature
for the 4-6 arcmin annulus. Contours are $\Delta \chi^2$= 2.6, 4.3 and 9.2,
corresponding to 68\%, 90\% and 99\% confidence levels for
two interesting parameters. The best-fit point is marked with a cross.
\label{contour}}
\end{figure}

\begin{figure}
\includegraphics[scale=0.35,angle=-90]{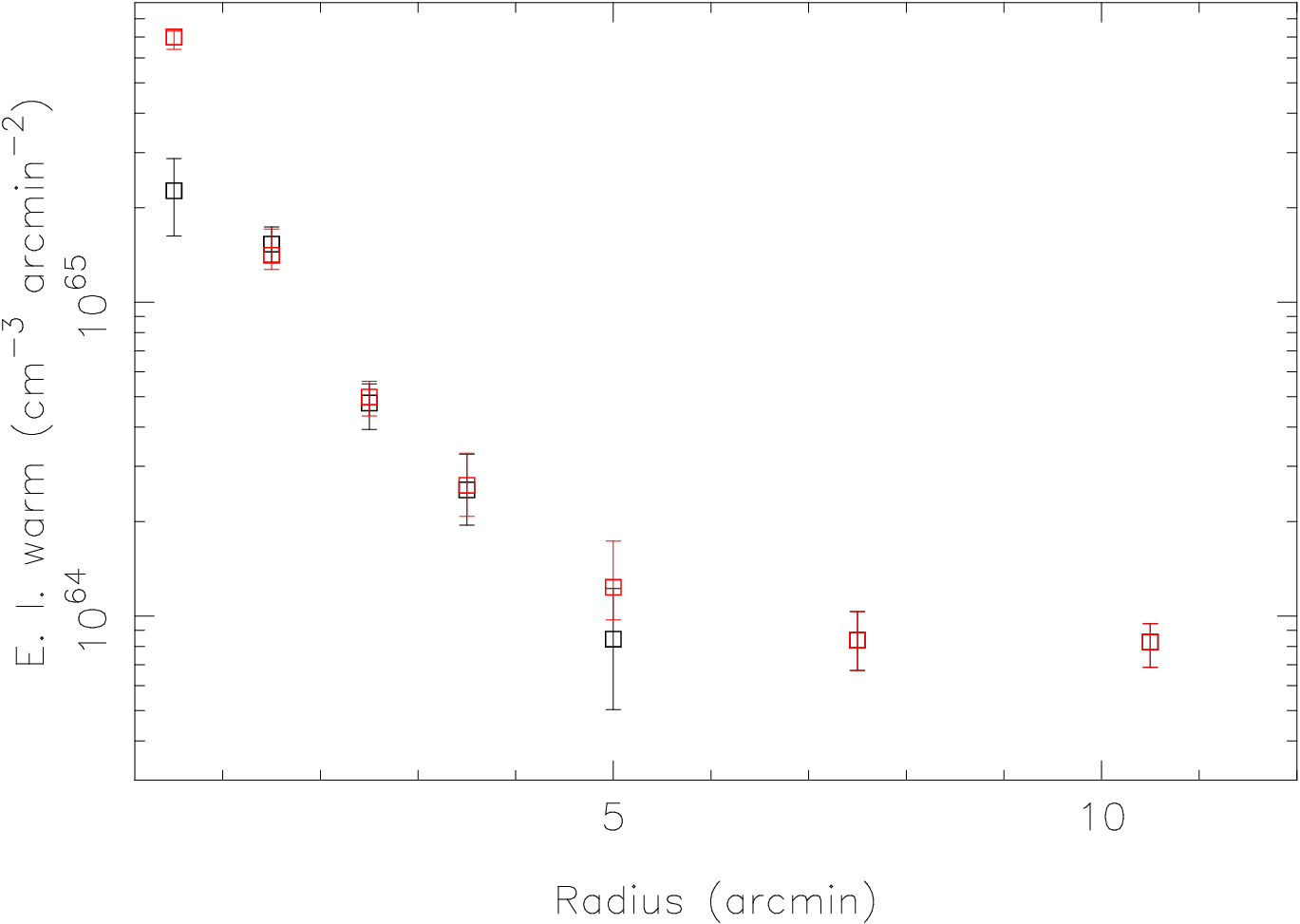}
\includegraphics[scale=0.35,angle=-90]{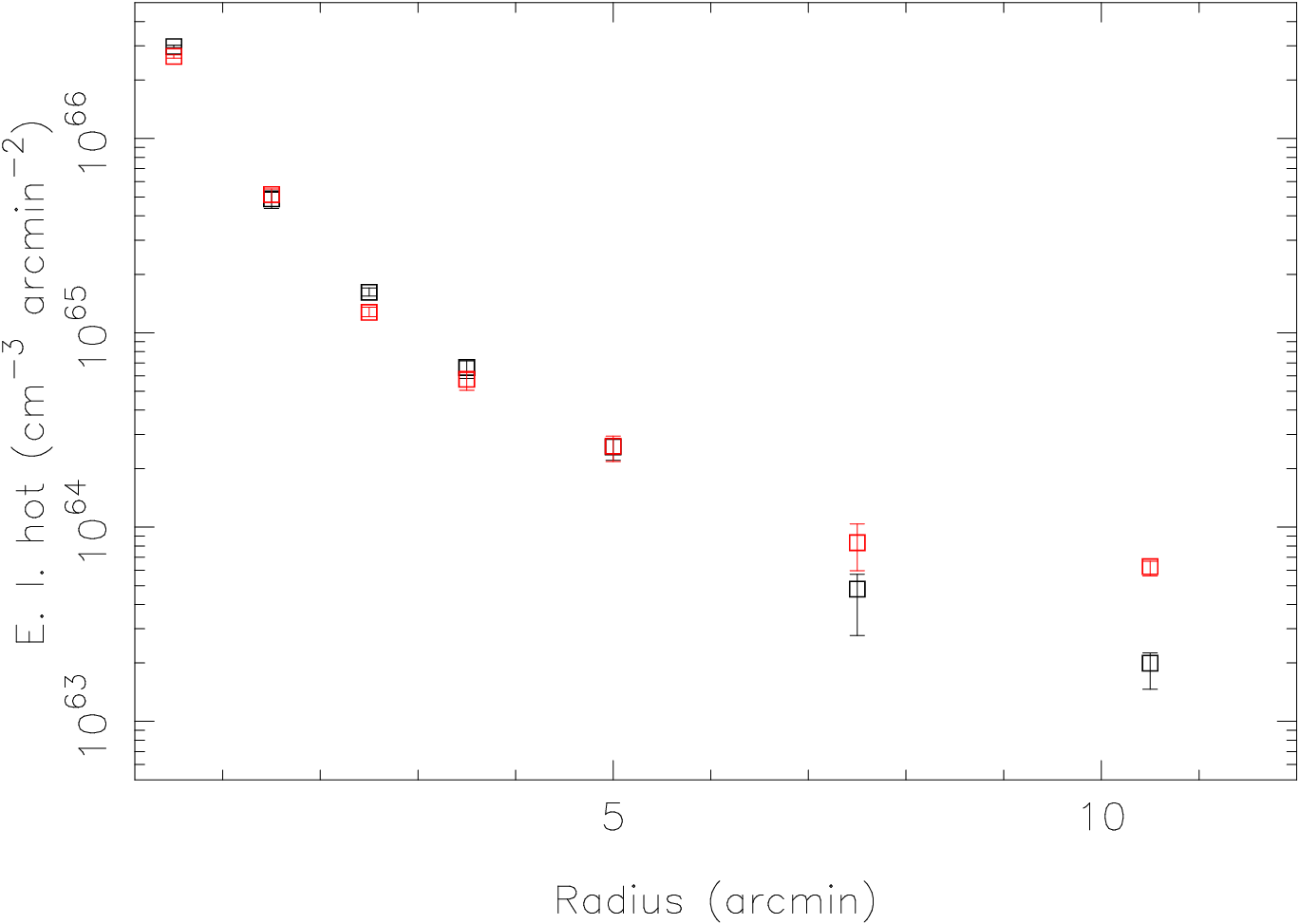}
\includegraphics[scale=0.35,angle=-90]{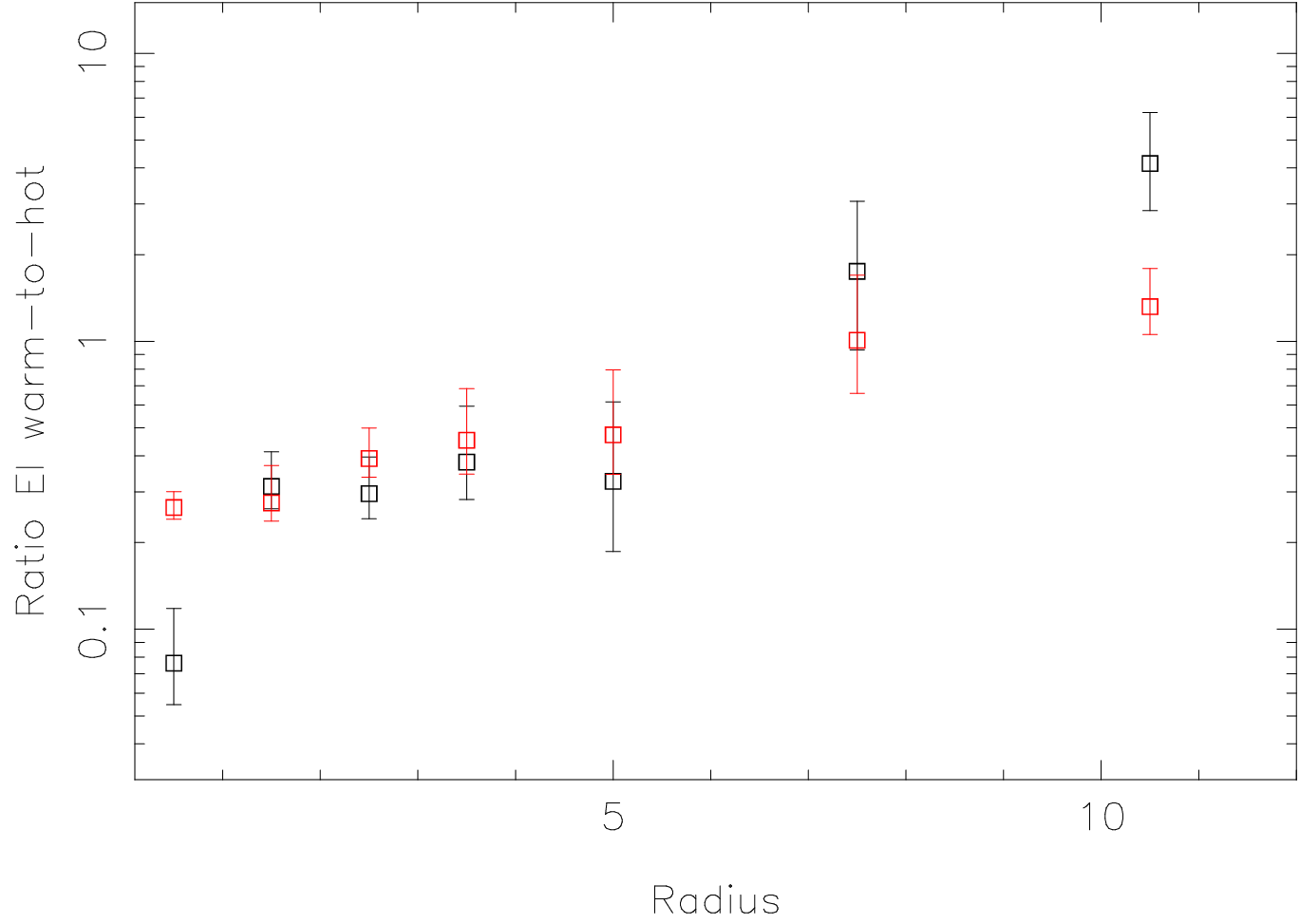}
\caption{
Emission integral per unit area of the warm and hot phases,
and their ratio, from the 2-temperature model.
In black is MOS data and in red pn data.
\label{eiwarm_hot}}
\end{figure}

First, we consider a 2-temperature model where both the hot ICM
temperature and the warm gas temperature are free parameters.
In the innermost two annuli the hot thermal component
features a $kT_{hot}/2$ sub-component, as described in section \ref{1t}.
This model constitutes a significant statistical improvement over the simple 
1-temperature model
(see Figure \ref{profiles}). The residuals of the 2-temperature model are now
consistent with the level of the calibration uncertainties, and are
shown in Figures \ref{A}(b) and \ref{B}(b) for two representative annuli.
The additional thermal component features a temperature of $kT_{warm}\sim$0.2-0.8 keV
and abundances of $\leq$0.1 solar (Figure \ref{warmprofiles}).
The abundance and temperature of the warm phase are correlated parameters. In Figure \ref{contour}
we show the contour plot of the two parameters for the region 4-6 arcmin, showing 
how 
the \xmm\/ data can constrain the physical properties of the warm plasma model.
Similar results apply to the other regions.
It is also clear, from the radial distribution of the emission integral ($I=\int n^2 dV$)
of the two phases, that the warm phase dominates 
at large radii, where the hot ICM emission becomes weak (Figure \ref{eiwarm_hot}).
\subsubsection{1-temperature model with central `cooling' gas plus power-law model \label{1tpo}}
\begin{figure}
\includegraphics[scale=0.5,angle=-90]{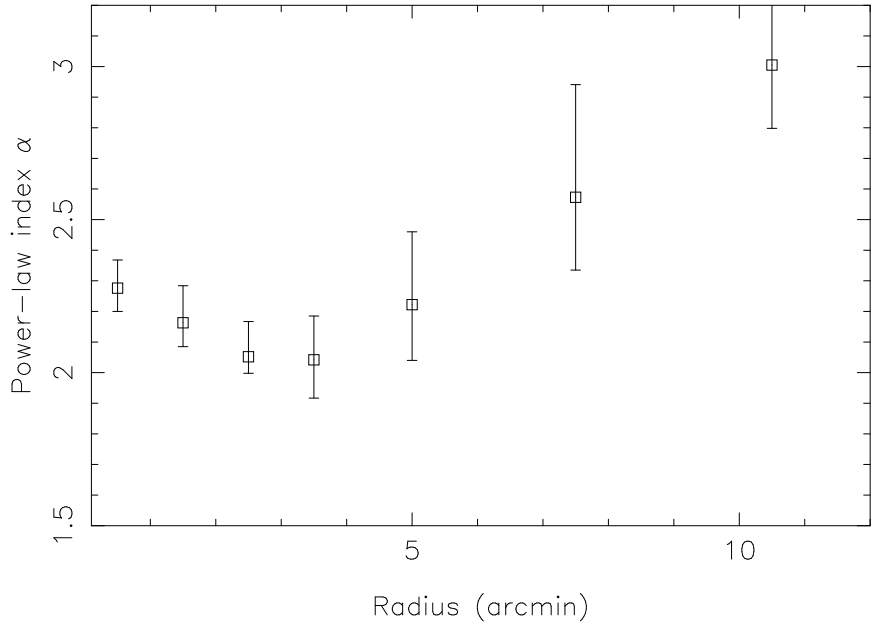}
\caption{
Power-law index profile of the
non-thermal model of section \ref{1tpo}.
%Normalization of the power-law model is in the
%customary XSPEC units; in black is MOS data and in red pn data.
\label{alphaprofiles}}
\end{figure}

We then consider a 1-temperature model plus a power-law component. 
This model is again an improvement over
the simple 1-temperature model, and the $\chi^2$ fit statistic indicates that
the 2-temperature model is only marginally preferred to this
non-thermal model (Figure \ref{profiles}).  The residuals of this non-thermal
model are shown in Figures \ref{A}(c) and \ref{B}(c) for two representative annuli.
In Figure \ref{alphaprofiles} we show the
radial profiles of the power-law index of the power-law component.

\section{Thermal interpretation of the soft excess emission \label{thermal}}

The 2-temperature model of section \ref{2t} provides  the most satisfactory fit to the 0.3-7 keV spectra.
In the following we  provide the physical interpretation of the thermal modelling of the
soft excess emission discussing two possible scenarios.
\subsection{Clumpy warm intra-cluster gas  {\it \`{a} la Cheng et al. (2005)} \label{cheng}}
\begin{figure}
\includegraphics[scale=0.5,angle=-90]{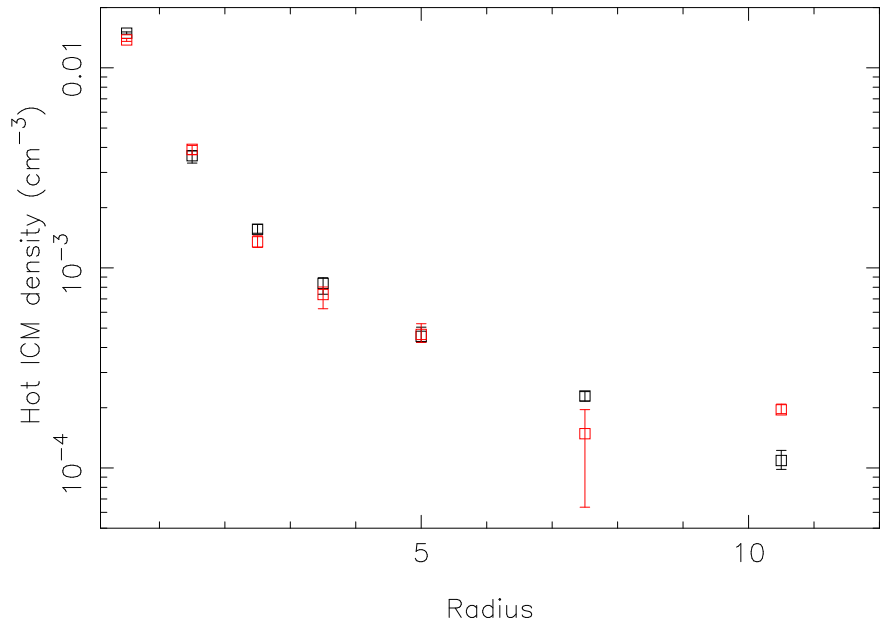}
\caption{Number density of the  hot gas from
the 2-temperature model.
In black is MOS data and in red pn data.
\label{density}}
\end{figure}

The warm gas responsible for the soft excess 
may be in a high-density ($\sim 10^{-2}-10^{-3}$ cm$^{-3}$) and clumpy 
intra-cluster phase with volume filling factor $f<<1$.
This is the scenario that results from the simulations of Cheng et al. (2005), 
in which a small fraction of the intra-cluster gas remains at low
temperature and high density during the accretion process.
This phase of the intra-cluster medium is
able to reproduce quantitatively the typical soft excess emission
detected in several clusters (e.g., Bonamente et al. 2002). 
%In this case, the mass estimates would be reduced by a factor of $\sqrt{f}$ with
%respect to those presented above in section \ref{warmicm}, and the
%cooling time is  reduced by a factor of $f$.
%Using $f=0.01$ as an example, the soft excess detected in this \xmm\/
%observation of \as\/ would track a mass of only $\sim$ 7\% of the hot gas mass.
%The cooling time would now be of less than 1 Gyr at all radii, and the
%replenishment of the warm gas becomes a dynamical phenomenon. 
We provide an estimate of the warm gas mass assuming that the warm gas is in
pressure equilibrium with the hot gas.
First, we deproject the emission integral $I_{hot}=\int n_{hot}^2 dV$ of the hot phase
(Figure \ref{eiwarm_hot}) to obtain the hot gas density (Figure \ref{density}).
Our results for the hot gas density are in very good agreement
with those from an earlier \xmm\/ observation by Kaastra et al. (2001).
The warm gas density is then calculated assuming pressure
equilibrium, $n_{warm} = (kT_{hot}/kT_{warm}) \cdot n_{hot}$, and it is in the
$10^{-2}-10^{-3}$ cm$^{-3}$ range.
Using the measured emission integral of the warm phase, defined as
$I_{warm}=\int n_{warm}^2\cdot f\cdot dV$, we estimate the volume filling
factor $f$ and hence the warm gas mass.
The volume filling factor is of a few percent ($f\simeq0.01-0.1$)
and the total warm gas mass is $M_{warm}\sim 3\times 10^{12}$ M$_{\odot}$
($3\pm 0.3 \times 10^{12}$ M$_{\odot}$ from MOS and $2.7\pm^{0.4}_{0.5} \times 10^{12}$ 
M$_{\odot}$ from pn data).
For comparison, the hot gas mass 
is
$M_{hot} \sim 1.2 \times 10^{13} M_{\odot}$ ($1.2\pm0.1 \times 10^{13} M_{\odot}$ from
MOS and $1.3\pm0.2 \times 10^{13} M_{\odot}$ from pn data).
Reiprich and Boheringer (2002) measure a total gravitational mass of
$M=1.8\times 10^{14} M_{\odot}$ out to the same radius of our analysis (0.9 Mpc),
resulting in a gas mass fraction of $f_{gas}\simeq0.07$.
If the excess emission follows the simple model presented in this section,
the warm gas will contribute to an additional $\sim$25\% of the
total baryonic mass.
The cooling time of the warm gas is shorter than the Hubble time at all radii,
and the replenishment of the warm gas becomes a dynamical phenomenon.

For comparison with the soft excess of the Coma cluster (Bonamente, Joy and Lieu 2003)
we also estimate the warm gas mass in the case of an intra-cluster phase with
volume filling factor $f=1.0$. In this case, the warm gas would not be
in pressure equilibrium with the hot gas and it would have a total mass of
$M_{warm} \sim 9\times 10^{12}  M_{\odot}$ ($9\pm2 \times 10^{12}  M_{\odot}$ from
MOS and $9\pm^2_1 \times 10^{12}  M_{\odot}$ from pn data).
This result is similar to that of Coma, where and $f=1.0$ warm gas yielded 
$M_w/M_h \sim 0.7$ (Bonamente, Joy and Lieu 2003).

\begin{figure}
\includegraphics[scale=0.5,angle=-90]{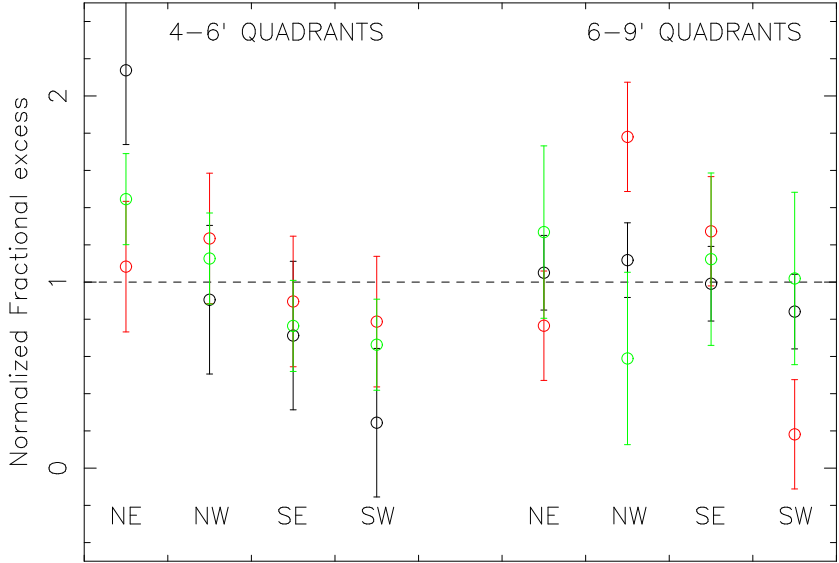}
\caption{Fractional excess $\eta$ in the 4 quadrants of annuli 4-6 and 6-9 arcmin, normalized
by the average fractional excess $<\eta>$ in that annulus. A value of  $\eta/<\eta>=1$
indicates that, in that sector, the soft excess is consistent with
the azimuthal average, and a value of $\eta/<\eta>=0$ indicates no excess.
In black and red the MOS1 and MOS2 data, and in green the pn data.
\label{D}}
\end{figure}

If the soft excess emission is indeed clumpy as described by the Cheng et al. (2005)
model, significant azimuthal variations of the excess could
be present.
In order to test the presence of azimuthal variations of the soft emitter,
we divided the two annuli with the highest quality data (4-6 and 6-9 arcmin)
into quadrants, and applied to them the analysis described in 
section \ref{2-7fit} to measure the fractional excess in the 0.3-1 keV band.
In Figure \ref{D} we plot the fractional excess detected in each quadrant, normalized by
the average fractional excess in that annulus.
At the resolution of our \xmm\/ data we cannot detect statistically 
significant azimuthal variations, although 
deviations at the level of $\pm$50\% of the average are allowed by the
data. A similar study of azimuthal variations of the soft excess in Coma
also yielded inconclusive results (Bonamente, Joy and Lieu 2003).

\subsection{Warm intergalactic gas \label{cen} {\it \`{a} la Cen and Ostriker (1999)}}

Alternatively, the warm gas may be located in a diffuse halo or 
filaments around the cluster, not in direct contact
with the hot ICM and extending beyond the cluster's virial radius. 

\begin{figure}
\includegraphics[scale=0.5,angle=-90]{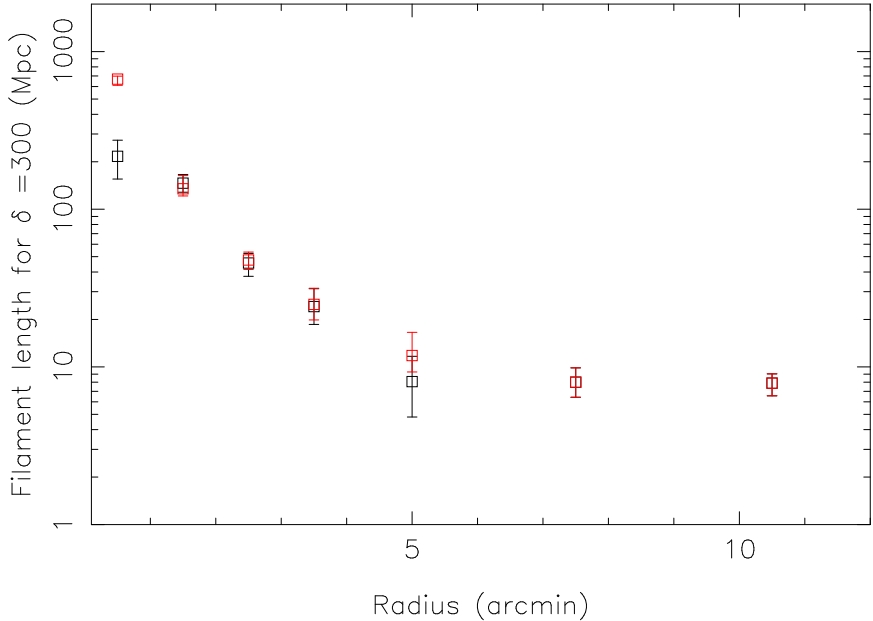}
\caption{
Length of filaments for the constant-density filament model of section \ref{cen}.
In black is MOS data and in red pn data. The large emission integral of the
warm gas (Figure \ref{eiwarm_hot}) causes the high estimates of the filament length.
\label{filament}}
\end{figure}

We assume that the detected warm emission resides in filaments of constant density $n_{fil}$. In this case,
the emission integral of each annulus becomes 
\begin{equation}
I_{warm}= \int n_{fil}^2 A dL
\end{equation}
where $dV=A dL$ is the volume of the filament, $A$ is the area of each annulus, 
and $L$ is the length of the filament along the
line of sight.
From the measured emission integral we cannot, at the same time, 
determine the density and the length of the filament.
If we assume that $n_{fil}=10^{-4}$ cm$^{-3}$, corresponding to a baryon overdensity of $\delta \sim 300$,
we can measure the filament lenght and mass 
implied by this model.
The filaments would be as long as an improbable few hundred Mpc (Figure \ref{filament}), 
and the total filament mass is $M_{fil} \sim 7.5 \times 10^{13} M_{\odot}$ 
($6.6\pm1.4 \times 10^{13} M_{\odot}$
from MOS and $8.4\pm1.3 \times 10^{13} M_{\odot}$ from pn data), resulting in a filament-to-hot
gas mass ratio of $\frac{M_{fil}}{M_{hot}} \sim 6$ ($5.5\pm1.3$ from MOS data and $6.4\pm1.4$ from 
pn data).
If the filaments are less dense, then the length and mass estimates would increase; in fact, $L\propto 
n_{fil}^{-2}$, and $\frac{M_{fil}}{M_{hot}} \propto n_{fil}^{-1}$ for a given emission
integral $I_{warm}$. These results are again similar to those
of the Coma cluster, where $\frac{M_{fil}}{M_{hot}} \sim$3 (Bonamente, Joy and Lieu 2003).
Similar models for the excess emission based on a diffuse halo of warm gas 
were  described in Kaastra et al. (2004b) and
Lieu and Mittaz (2004).
One can derive similar results for the filament mass of this scenario assuming that
the filaments have a given size, instead of a given density as assumed above.
In this case, the large emission integral of the warm gas measured at small radii
(Figure \ref{eiwarm_hot}) implies denser filaments towards the central cluster regions.

Mittaz et al. (2004) investigated whether the
soft excess emission could be quantitatively explained by the large-scale filaments
seen in the simulations of Cen and Ostriker (1999), Cen et al. (2001) and Dav\'{e} et al. (2001).
The emission from filaments around the simulated clusters was a few orders of magnitude
below the level needed to explain the typical soft excess emission as, e.g., that
of \as.
We therefore alert the reader that 
those simulations are not in quantitative agreement with the soft excess measured
in \as\ and in other clusters.

\begin{figure}
\includegraphics[scale=0.5,angle=-90]{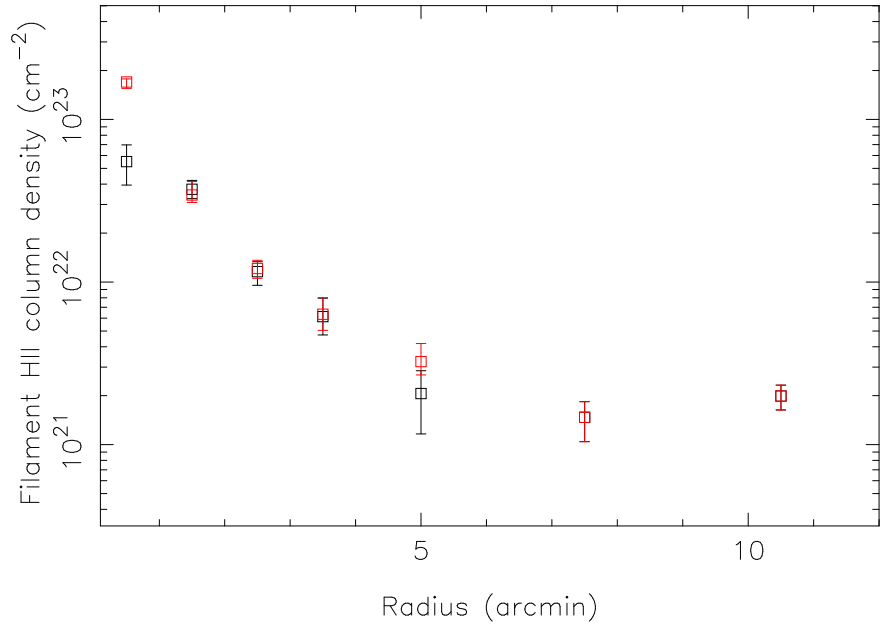}
\caption{Column density of HII according to the warm filament model
of section \ref{cen}.
\label{filamentcolumnden}}
\end{figure}

Other simulations by Fang et al. (2002) also find that a large fraction of baryons is in
a warm filamentary phase of the intergalactic medium. If the excess emission of \as\/ is due
to the filaments  model described in this section, the excess implies HII column densities
$N_{HII} > 10^{21}$ cm$^{-2}$, as shown in Figure \ref{filamentcolumnden}.
The associated OVII and OVIII column density depends on the exact ionization structure of
the gas: assuming a ionization fraction of $\sim 10\%$, valid for a gas in
collisional equilibrium at $kT \simeq 0.3$ keV (e.g., Fang et al. 2002), and an average
metal abundance of 5\% solar (see section \ref{complexmodels}), 
the OVII and OVIII column densities implied by the
excess emission of \as\/ exceed $N_{OVII/OVIII} > 2\times 10^{15}$ cm$^{-2}$.~\footnote{This
calculation assumes an oxygen abundance of $8.5 \times 10^{-4}$ relative to the
hydrogen number density, and oscillator frequencies of 0.696 and 0.416 for the
OVII and OVIII absorption lines (e.g., Fang et al. 2002).}
This result is qualitatively consistent only with the brightest 
regions of the Fang et al. (2002) simulations
(see Figures 4 and 5 of Fang et al. 2002).

\subsection{Discussion of thermal interpretation}

The first thermal scenario (section 
and \ref{cheng}) is preferred to the second (section \ref{cen}),
given that the calculations of Mittaz et al. (2004) show
no quantitative agreement between the measured soft excess and
the simulation predictions.
In the case of a clumpy intra-cluster  gas {\it \`{a} la Cheng et al. (2005)}
a shorter-than-Hubble cooling time is a inherent feature of the model,
and the soft excess emission is a dynamical phenomenon.
The filaments {\it \`{a} la Cen and Ostriker (1999)} discussed in section \ref{cen} 
would be as long as several 100 Mpc, or even in excess of 1 Gpc if the 
filaments are less dense than $n=10^{-4}$ cm$^{-3}$, in disagreement
with the current observational and theoretical understanding of large-scale
structures.

The total baryon content of clusters is a useful probe of cosmology.
The thermal interpretation of the soft excess implies a warm gas mass that may be comparable
to that of the hot ICM, and therefore significant when evaluating the
baryon mass in clusters. 
Allen et
al. (2004) analyze a sample of clusters at z$<$0.9 and find that the
baryonic fraction in that sample is f$\sim$0.12. In calculating the total
baryonic mass, they account for the X-ray emitting hot ICM and the
luminous matter, and use a bias factor $b$ that accounts for the fact
that the baryon fraction in clusters, as seen in simulations (Eke et
al. 1998), is lower than for the universe as a whole by about 15\%.
They give constraints on $\Omega_{\Lambda}$ and $\Omega_m$ that
are in agreement with other measurements (SN1a and CMB measurements).
The soft excess emission in \as\/ and in other clusters (e.g.,
Bonamente et al. 2002) may be indicative of a massive warm
baryonic component in clusters. Given our mass estimates for
the warm baryons provided in section \ref{thermal}, we envisage two
possibilities that conciliate the thermal origin of the soft excess with
the results of Allen et al. (2004): the warm baryons have a mass of $\leq$15\%
of the hot gas mass or the
warm gas, being in structures that are not virialized (e.g., the model
in section \ref{cen}), should not be counted as "cluster" baryons.

\section{Non-thermal interpretation of the soft excess emission \label{nonthermal}}

Alternatively, the excess emission may be of non-thermal nature. In this section
we investigate the physical interpretation of the non-thermal modelling of
section \ref{complexmodels}.

\subsection{IC scattering of relativistic e$^-$ off the CMB}

A population of relativistic electrons in the cluster would scatter the
cosmic microwave background (CMB), and give rise to soft X-ray radiation via
an inverse Compton scattering (e.g., Huang 1997; Sarazin and Lieu 1998; Lieu et al. 1999).
A relativistic electron with a Lorentz factor $\gamma$ will up-scatter
a  CMB photon  to an energy of
\begin{equation}
E= 75 \times \left(\frac{\gamma}{300} \right)^{2} \; \rm eV.
\end{equation} 
Electrons with $\gamma=600-1500$ are needed to generate soft excess photons
in the 0.3-2 keV band.

Relativistic electrons in clusters may be generated by diffusive shock acceleration.
The acceleration produces a power-law distributed population of electrons which,
upon scattering with the CMB photons, creates power-law radiation spectrum
(Ribicki and Lightman 1979). The relationship between the two power-law indices is
\begin{equation}
\mu=-1+ \alpha \cdot 2
\end{equation}
where $\mu$ is index of the differential number electron distribution, and
$\alpha$ is the differential photon number index.
For Galactic cosmic ray electrons it is typically observed an index $\mu \sim 2.7$,
corresponding to $\alpha \sim 1.85$.
Relativistic electrons undergo Coulomb losses at low-energies
and radiative losses at higher energies. In the $\gamma=600-1500$
regime (where $\gamma$ is the electron Lorents factor) 
and for typical cluster conditions, radiative losses will steepen 
the electron spectrum in a few $10^{9}$ years
since acceleration (Lieu et al. 1999; Ip and Axford 1985).  
The results of Figure \ref{alphaprofiles} show a power-law index $\alpha \geq$2. 
This indicates that, if the soft excess is of 
cosmic ray origin, the electron population was injected in the
cluster environment a few $10^9$ years ago.

\begin{figure}
\includegraphics[scale=0.5,angle=-90]{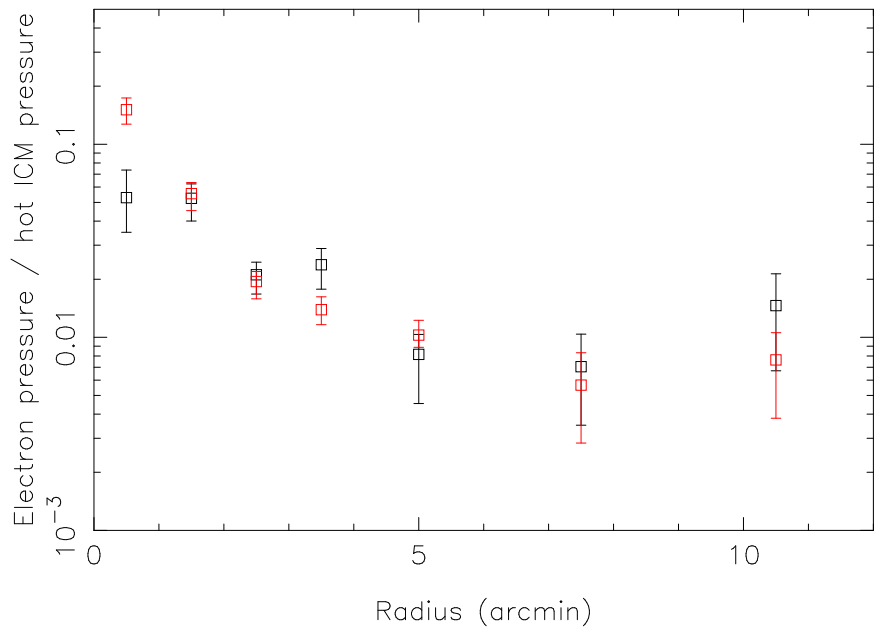}
\caption{Pressure ratio of relativistic electrons and hot ICM plasma
in the non-themal model.
In black is MOS data and in red pn data.
\label{pressure}}
\end{figure}

The  modelling of section \ref{complexmodels} yields the luminosity
of the power-law component which can then be used to determine the energy of
the underlying electron population (Lieu et al. 1999), via
\begin{equation}
E_{e}=8 \times 10^{61} L_{42} \left( \frac{3-\mu}{2-\mu}\right) \frac{\gamma_{max}^{2-\mu}-\gamma_{min}^{2-\mu}}
{\gamma_{max}^{3-\mu}-\gamma_{min}^{3-\mu}} \;\;\rm ergs,
\end{equation}
where $L_{42}$ is the luminosity of the non-thermal radiation in units of $10^{42}$ erg s$^{-1}$,
$\gamma_{min}$ and $\gamma_{max}$ the Lorentz factors corresponding to the 
boundaries of the \xmm~ data
(0.3-7 keV).
From this, one can calculate the electron pressure ($P=E_{e}/3 V$)
and compare it with the hot ICM pressure. The resuls are shown in Figure \ref{pressure},
which indicate that the non-thermal electron pressure is a few \% of the thermal pressure at all radii.

\subsection{Discussion of the non-thermal interpretation}

If the soft excess is of non-thermal nature, relativistic electrons must be present throughout the 
central 1 Mpc radius of \as. The cluster has no known signature of non-thermal activity (e.g., no
detected radio halo or strong radio sources). It is possible, however, that
non-thermal activity may have occurred in the past. The steeper-than-Galactic
power law indices indicate an aging of the non-thermal population.

The presence of relativistic particles at a distance of $\sim$1 Mpc from the cluster center
is challenging. In fact, the Bohm diffusion time for relativistic electrons
exceeds a Hubble time in the typical cluster environment (Volk, Aharonian and Breitschwerdt 1996).
The widespread presence of relativistic particles would require several
sites of local acceleration, 
or alternative means of transport from the acceleration site to $\sim$Mpc distances.
The acceleration could be provided by several shocks (e.g., accretion shocks) 
or, alternatively, AGN jets, of which we have no evidence at present.
Diffuse shocks accelerate ions as well as electrons (Bell 1978a,b; Axford,
Leer and Skadron 1977; Blandford and Ostriker 1978). The ions, due to the higher mass,
do not emit effectively, but will carry the
bulk of the energy, exceeding the electron energy  by a factor of
$\sim20-50$. This in turn implies that the estimates of Figure \ref{pressure}
will have to be increased by the same factor, with the relativistic particles
now in approximate energy equipartition with the hot ICM pressure. This result was already noted in
the Coma cluster (Lieu et al. 1999) and in our previous study of
\as~ with  ROSAT data (Bonamente, Lieu and Mittaz 2001). With the current \xmm~ data 
we have provided more precise estimates of the non-thermal pressure in \as.
The presence of relativistic ions can only be avoided if the electrons are accelerated
by an alternative means that do not accelerate the ions, such as AGN electron/positron jets.

\section{Origin of the soft excess emission of \as \label{interpretation}}

The soft excess emission of \as\/ is of genuine celestial origin.
Knowledge of the \xmm\/ detectors and direct use of the calibration
source \pks\/ (section \ref{calibration}) showed that the emission is not a detector artifact.
Moreover, the emission cannot be caused by wrong assumptions on the 
HI Galactic absorbing column. In order to explain the excess emission, the
Galactic column density is required to be more than two times {\it lower} than
the measured Galactic value ($1.8\pm0.1 \times 10^{20}$ cm$^{-2}$) and even lower than
the Lockman Hole value ($\sim 7\times 10^{19}$ cm$^{-2}$) for certain annuli.
Bregman et al. (2003) showed that anisotropies in the distribution of the
HI column density towards galaxy clusters may affect the transmission
of extragalactic soft X-ray radiation. Such effect, however, is only relevant at
energies $E<0.3$ keV (e.g., Figures 1 and 9 of Bregman et al. 2003), and
it has no influence on the soft excess detected in this paper at
$E=0.3-2$ keV with the \xmm\/ detectors.

The thermal and non-thermal models examined in this paper
constrain the nature of the soft excess emission from \as.
The formal $\chi^2$ goodness-of-fit statistic does not provide a definitive
indication of the origin of the soft emission.
The thermal scenario is only marginally favored over the non-thermal
(Figure \ref{profiles}), similarly to
the case of the Coma cluster (Bonamente, Joy and Lieu 2003).
The large extent of the detected excess emission
($\sim$ 1 Mpc radius) renders unlikely a purely non-thermal interpretation
of the excess, due to the long diffusion times of relativistic
particles which cannot easily diffuse throughout the
entire cluster volume (see section \ref{nonthermal}).

The \xmm\/ data of \as\/ indicate that
warm gas located within the virial radius provides the
bulk of the soft excess radiation.
The gas may be in a 
clumpy phase of the ICM (section \ref{cheng}) as advocated
by the simulations of Cheng et al. (2004), with a mass
of 25\% of the hot gas mass.
Although we do not confirm the direct detection of OVII and OVIII
emission lines associated with the excess emission, the warm gas emission is
consistent with a metal abundance of a few \% solar.
The soft emission is unlikely to originate from large-scale
filaments {\it \`{a} la Cen and Ostriker (1999)}, which  would encompass a total 
mass in excess of that of the hot ICM.

It is possible
that a fraction of the soft excess, especially in the central regions, is of
non-thermal nature (section \ref{nonthermal}). Although there is no evidence at present
of non-thermal activity in \as, past AGN or diffusive shock acceleration may have fostered the
presence of relativistic electrons in the cluster environment.
Our calculations show that the relativistic particles would be near 
or below pressure equipartition with the hot gas.

\section{Comparison with earlier observations of \as \label{comparison}}
The initial report of soft excess in \as\/ was based on low resolution
\rosat\/ PSPC data (Bonamente, Lieu and Mittaz 2001). The \rosat\/ data were able
to detect soft excess emission out to a radius of $\sim$9 arcmin.
The mass estimates for the warm intra-cluster gas model had significantly
larger errors, but are in agreement with the present \xmm\/ results.
The non-thermal fit results based on these earlier data have a power-law
index that is significantly higher than that of the present analysis (see
Table 2 of Bonamente, Lieu and Mittaz 2001). The reason is likely attributable
to the narrower \rosat\/ band, and it is the cause for the higher non-thermal
perssure estimates we obtained based on that observation.

The first \xmm\/ results on \as\/ were based on an earlier $\sim 30$ ks
observation, and are described in Kaastra et al. (2001).
In that paper no attempt was made to detect or measure the soft excess,
but the authors focused on the properties of the hot gas.
Our estimates of the hot gas temperature, abundance and density are in good 
agreement with theirs (see Figures 1--3 of Kaastra et al. 2001).

Kaastra et al. (2003) analyzed the same \xmm\/ observation as
Kaastra et al. (2001) in order to investigate the  soft excess emission.
They report the presence of soft excess emission out to a radius of
$\sim$12 arcmin, in agreement with our present results. Their thermal
and non-thermal models for the excess emission also indicate that the
two models have similar goodness of fit in the central regions, in agreement
with our present findings.
The backround subtraction technique they employ, however, differs from the one used in
this paper in that  we now use a local background.
The different background subtraction is the reason for a variance with our results
at large radii, e.g., their detection of OVII emission lines associated with the
soft excess emitter is not confirmed in our analysis (see section \ref{background}).

The same authors provide an estimate of the warm gas mass in Kaastra et al. (2004b), based
on the same earlier \xmm\/ observation. Their warm gas estimate follows a model
reminiscent of our warm intergalactic filaments {\it \`{a} la Cen and Ostriker (1999)}.
They assume a fiducial radius for the volume of the emission and estimate a constant
density for the warm gas, and find a warm gas mass of $\sim 6\times 10^{14} M_{\odot}$,
higher than our corresponding estimate by a factor of $\sim$10.
The difference can be explained by the simplifying assumptions of their model.

The properties of the central `cooling' gas of \as\/ were also studied in detail
by Kaastra et al. (2004a). They find that cooling gas is significantly present only 
at radii $<$2 arcmin. Our best-fit hot ICM temperatures are again in agreement with those
of Kaastra et al. (2004a).

\section{Conclusions}
%In this paper we have analyzed and interpreted the highest quality data available to date for a soft
%excess cluster.
%First, we showed that the soft excess emission is strongly detected 
%throughout a $\sim 1$ Mpc radius region, confirming earlier detections based on
%lower S/N data (Kaastra et al. 2003, Bonamente et al. 2001).
%We then investigate several thermal and non-thermal models of the emission, and provide the 
%resulting mass and energy budgets.
%These models provide useful constraints on the origin of the soft excess emission,
%and enable us to conlcude that the soft excess of \as\ is predominantly
%of thermal origin, that the gas very likely located within the virial radius and that
%part of the excess emission can also be of non-thermal origin.
In this paper we have analyzed a high S/N \xmm\/ observation
of \as and provided  a wide range of models for the 
soft excess emission.
We employed a different method of background subtraction than the earlier
\xmm\/ observations of Kaastra et al. (2003) which accounts
for spatial variations of the soft X-ray background.

First, we showed that soft excess emission is strongly detected throughout a $\sim 1$ Mpc
radius region, confirming earlier detections based on lower S/N data (Kaastra et al. 2003;
Bonamente, Lieu and Mittaz 2001). We ruled out the possibility that the excess emission
is caused by peculiarities in the absorbing column of Galactic gas, or that the emission
is an artifact of the \xmm\/ detectors.

We then investigated several thermal and non-thermal models in search of
the origin of the soft excess emission. The thermal models assume the presence of
warm ($\sim 10^6-10^7$ K) gas either intermixed with the hot gas, or in diffuse
structures extending beyond the cluster virial radius. The non-thermal model requires the
widespread presence of relativistic electrons with Lorentz factor $\sim1000$.

We calculated the density and mass budget for the thermal models and the pressure
budget for the non-thermal model, and concluded that the most likely interpretation
for the soft excess emission of \as\/ is warm gas located within the virial radius
of the cluster, which contributes an additional 25\% of the total cluster baryonic mass.

%\begin{figure}
%\includegraphics[scale=0.5,angle=-90]{f12.eps}
%\caption{Cooling time of the warm intra-cluster gas.
%In black is MOS data and in red pn data.
% \label{cooling_time}}
%\end{figure}
                                                                                         
\end{document}